\documentclass[reprint,amsmath,amssymb,aps,prl,superscriptaddress]{revtex4-2}

\usepackage{graphicx}
\usepackage{dcolumn}
\usepackage{bm}
\usepackage{braket}
\usepackage{color}
\usepackage{pstricks,pst-grad,color}
\usepackage{appendix}
\usepackage{here}
\usepackage{url}
\usepackage{hyperref}
\usepackage{comment}
\usepackage{mathtools}

\newcommand{\mr}[1]{\mathrm{#1}}
\newcommand{\tr}[1]{\mathrm{Tr}\Bigl[#1\Bigr]}

\usepackage{tikz}
\usetikzlibrary{shapes,calc}
\usetikzlibrary{shapes.geometric,arrows.meta,decorations.markings}

\begin{document}
\preprint{APS}

\title{Machine-learning-assisted construction of appropriate rotating frame}
\author{Yoshihiro Michishita}
\email{yoshihiro.michishita@riken.jp}
 \affiliation{RIKEN Center for Emergent Matter Science (CEMS), Wako, Saitama 351-0198, Japan}

\date{\today}

\begin{abstract}
Machine learning with neural networks is now becoming a more and more powerful tool for various tasks, such as natural language processing, image recognition, winning the game, and even for the issues of physics. 
Although there are many studies on the application of machine learning to numerical calculation and the assistance of experimental detection, the methods of applying machine learning to find the analytical method are poorly studied.
In this letter, we propose methods to use machine learning to find the analytical methods. We demonstrate that the recurrent neural networks can ``derive'' the Floquet-Magnus expansion just by inputting the time-periodic Hamiltonian to the neural networks, and derive the appropriate rotating frame in the periodically-driven system. We also argue that this method is also applicable to finding other theoretical frameworks in other systems. 
\end{abstract}

\maketitle

{\it{Introduction.}} --
Machine learning with neural networks (NN) is now becoming a powerful tool for various tasks, such as natural language processing\cite{8416973}, image recognition\cite{NIPS2012_c399862d, 7780459, https://doi.org/10.48550/arxiv.1901.04407}, and winning the game\cite{Silver2017}. As for the issues of physics, machine learning can be used for phase detection\cite{PhysRevB.94.195105, Carrasquilla2017, vanNieuwenburg2017, PhysRevE.95.062122, PhysRevB.97.205110, PhysRevLett.120.176401}, solving the equilibrium state\cite{PhysRevB.96.205152, doi:10.1126/science.aag2302, Carleo2018} or steady state\cite{PhysRevB.99.214306}, materials informatics\cite{Brockherde2017, doi:10.1126/science.abj6511, Guo2022, Pederson2022} and noise reduction of the experimental measurement results\cite{Tsukamoto_2022}. While experimental detection and numerical calculation with the aid of machine learning are now making great progress, one might have a natural question: 

{\it Can we develop theoretical analysis methods with the aid of machine learning?}

Before tackling this question, let us consider how we ourselves have developed the theoretical analysis methods so far. One of the central techniques is scale separation, which leads to the reduction and the perturbation theory. A typical example is a derivation of the Langevin equation, in which, by utilizing the time-scale separation between the microparticles' motion and the Brownian particles' motion, we perform the reduction of the degrees of freedom of the microparticles and get the stochastic equation of motion.\cite{https://doi.org/10.48550/arxiv.cond-mat/0412296} Other examples are the dimensional reduction of nonlinear dynamical systems\cite{Haken1987}, the derivation of the Heisenberg model from the Hubbard model in half-filling and large interaction limit\cite{Heisenberg1928}, the renormalization group methods\cite{RevModPhys.47.773}, and so on.
However, in general, it is a non-trivial problem to find and separate the fast process and the slow process of the system, and we have to perform an appropriate unitary transformation or projection and get the frame in which the scale separation is apparent. \cite{PhysRev.90.297, PhysRev.149.491, doi:10.1063/1.447055, PhysRevLett.121.026805, Imbrie2016, PhysRevLett.125.180602, PhysRevLett.126.153603, PhysRevResearch.4.023194}

In periodically-driven systems, we introduce an appropriate rotating frame (RF) associated with the time-periodic unitary transformation and separate fast and slow modes. 
It is known that, under high-frequency driving where the frequency is larger enough than the energy scale of the system, the system stays the Floquet prethermalized before going to the infinite temperature state\cite{PhysRevLett.116.120401, KUWAHARA201696, PhysRevX.9.021027}, and we can engineer the desired state in the Floquet prethermalized state.\cite{doi:10.1080/00018732.2015.1055918, RevModPhys.89.011004, doi:10.1146/annurev-conmatphys-031218-013423, PhysRevLett.102.100403, Jotzu2014, PhysRevLett.116.205301, McIver2020, PhysRevLett.114.140401}
In the high-frequency regime, we can construct an appropriate RF with high-frequency expansion.\cite{PhysRevLett.128.050604} In such an appropriate RF, the effective static Hamiltonian describes the Floquet prethermalized state, and the dressed driving term describes the heating rate.\cite{PhysRevB.93.144307, PhysRevLett.128.050604}
Therefore, finding an appropriate unitary transformation or projection and a scale separation is highly beneficial, while it is usually difficult.

In this letter, we propose a method to search for the appropriate frame with desirable properties using machine learning techniques. Our method has the advantage that it is enough to set the desirable properties as a loss function and, thus, should be versatile. Moreover, our method has a possibility of finding ``implicit'' scale separation, which is difficult for the human to find, and the related appropriate frame.
 As a first step, in this letter, we demonstrate that machine learning can derive the appropriate rotating frame when there is an apparent scale separation. As concrete systems, we focus on a periodically-driven system under high-frequency driving and its appropriate rotating frame.

We first show the concrete procedure to construct the RF in periodically-driven systems with the aid of machine learning and demonstrate it for a simple model. Then, we argue how we can generalize the above procedure to derive an appropriate unitary transformation under which we can easily solve the issues of physics.
Our results show that machine learning is a powerful tool also in exploring theoretical analysis methods.

\

\begin{figure}
    \centering
    \includegraphics[width=0.98\linewidth]{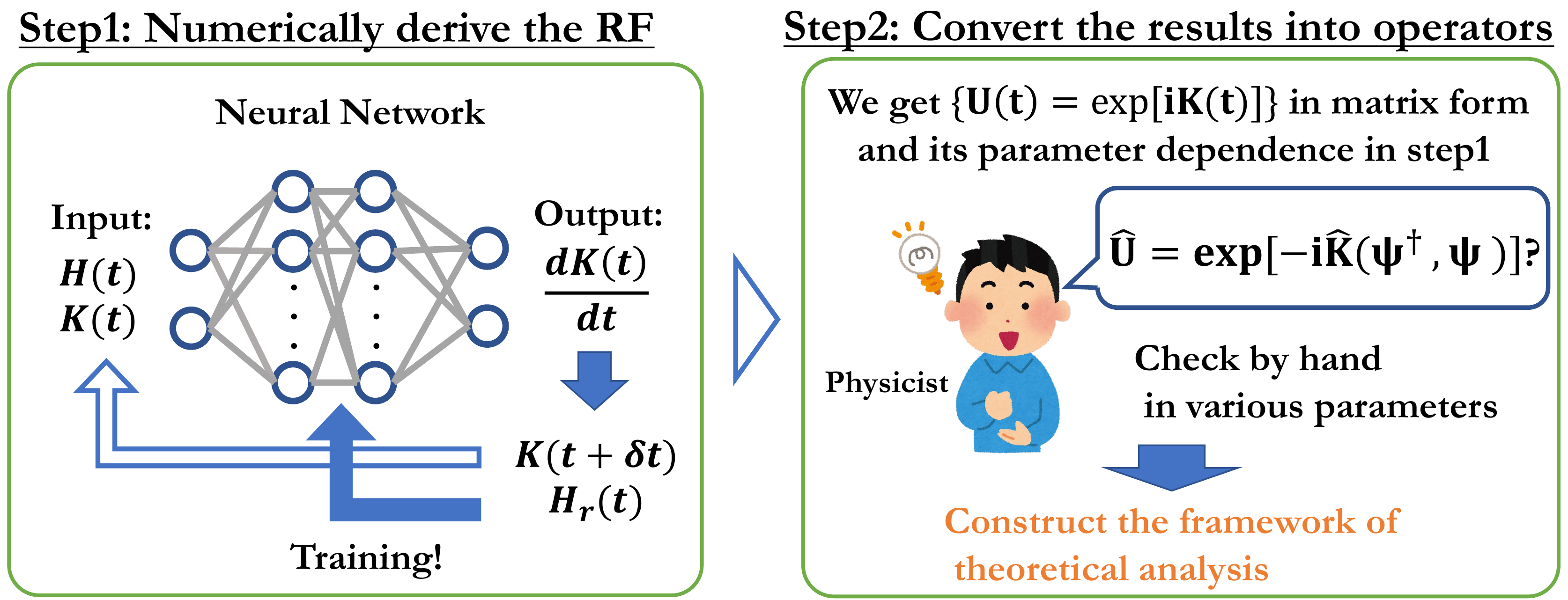}
    \caption{Schematic picture of machine-learning-assisted construction of appropriate rotating frame.}
    \label{fig:notionRF}
\end{figure}

{\it{Concrete procedure to construct the rotating frame with the aid of machine learning.}} --
The procedures to derive the RF are summarized in Figure.~\ref{fig:notionRF}, and it can be divided into two steps.
Below, we first review the Floquet theory and the RF shortly, and then explain each step.

We consider the time-dependent Hamiltonian as
\begin{align}
    \hat{H}(t) = \hat{H}_0 + \hat{V}(t),
\end{align}
where $\hat{H}_0$ is the system Hamiltonian without the driving and $\hat{V}(t)$ is the driving term which depends on time. Under the time-dependent unitary transformation, the Schr\"{o}dinger equation  $i\frac{d}{dt}\ket{\psi(t)} = \hat{H}(t)\ket{\psi(t)}$ can be written as
\begin{eqnarray}
    i\frac{d}{dt}\ket{\tilde{\psi}(t)} &\equiv& i\frac{d}{dt} \hat{U}(t) \ket{\psi(t)}
    =\hat{H}_r(t) \ket{\tilde{\psi}(t)},\\
    \hat{U}(t) &=& \exp\Bigl[i \hat{K}(t)\Bigr],\\
    \hat{H}_r(t) &\equiv & \hat{U}(t) \hat{H}(t)\hat{U}^\dagger(t) + i \bigl(\frac{d}{dt}\hat{U}(t)\bigr)\hat{U}^\dagger(t)\\
    &=&\hat{U}(t)\Bigl(\hat{H}(t) - i \frac{d}{dt}\Bigr)\hat{U}^\dagger(t),\label{Hr}
\end{eqnarray}
where we call $\hat{H}_r(t)$ as the dressed Hamiltonian, which describes the dynamics of $\ket{\tilde{\psi}(t)}$ from the reference frame of $\hat{U}(t)$.

When we focus on the periodically-driven systems, where $\hat{H}(t)=\hat{H}(t+T)$ and $\hat{V}(t)=\hat{V}(t+T)$ are satisfied, it is known that there is the periodic time-dependent unitary transformation $\hat{U}_F(t) = \hat{U}_F(t+T)$, which make the dressed Hamiltonian time-independent, that is, $\hat{H}_r(t) = \hat{H}_F$. We call $\hat{H}_F$ Floquet Hamiltonian,  $\ket{\tilde{\psi}(t)} = \hat{U}_F(t)\ket{\psi(t)}$, whose dynamics is described by the Floquet Hamiltonian, the exact rotating frame, and $\hat{K}(t)$ the micromotion operator or the kick operator. 
In the high-frequency regime, we can construct this exact rotating frame perturbatively with high-frequency expansion such as the Floquet-Magnus expansion\cite{PhysRevLett.116.120401, KUWAHARA201696} and van-Vleck expansion\cite{doi:10.1080/00018732.2015.1055918, Eckardt_2015}. If we calculate up to the finite order $\mathcal{O}(1/\Omega^n)$, the dressed Hamiltonian slightly depends on time by $\mathcal{O}(1/\Omega^{n+1})$.
Fortunately, when we calculate the heating rate of the system under the high-frequent driving, it is more beneficial to calculate the high-frequency expansion up to the second or third order \cite{PhysRevLett.128.050604} and we call this benefitial RF an appropriate RF.

In the following, supposing that the physicists in the world do not know the high-frequency expansion methods and the appropriate RF but want to derive the appropriate time-periodic unitary transformation in which the dressed Hamiltonian slightly depends on time, we ``derive'' the high-frequency expansion and the appropriate RF with the aid of machine learning.

In the first step (step 1), we numerically derive the RF with the recurrent neural networks (RNN) \cite{Rumelhart1986, ELMAN1990179, WERBOS1988339}. We input the model Hamiltonian $H(t)$ and the kick operator $K(t)$, which is calculated by the NN, and get the output as the time derivative of the kick operator $K'(t)$. Then, we can calculate the kick operator $K(t+\delta t) = K(t) + K'(t)\delta t$ and the dressed Hamiltonian $H_r(t) = U(t)H(t)U^{\dagger}(t) -iU(t) \{U(t+\delta t) - U(t)\}/\delta t$.
We set the boundary condition $K(t=0) = 0$; therefore, $K(t)$ is given at $t=0$, and we can input it into the NN.
We note that, in this letter, we use $\hat{O}$ when it describes the operator form and $O$ when it describes the matrix form numerically.
Because we want to get the RF in which the time-dependence of the dressed Hamiltonian is small, we set the loss function at the time step $t$ of learning as,
\begin{eqnarray}
    l(t) = \sum_{n} \gamma^n ||H_r(t) - H_r(t-n\delta t)||,
\end{eqnarray}
where $||M|| = \tr{M^2}$ and $\gamma$ is the discount rate, which is usually introduced in the reinforcement learning and $0<\gamma \leq 1$. We note that we can consider step1 in Figure.\ref{fig:notionRF} as a reinforcement learning in which NN as an actor decides the rotating frame at the next time step in each turn and the cost $l(t)$ is given depending on the action. After that, because we now have the kick operator at the next time step $K(t+\delta t)$, we repeatedly do the same procedure until the time step reaches $t=T$, and calculate the total loss function as
\begin{eqnarray}
    l = \sum_{t} l(t) \delta t + \epsilon||\sum_{t} K'(t)\delta t||,\label{loss}
\end{eqnarray}
where $\mathrm{N_{t}} = T/\delta t$ is the total time step, and update the parameters of NN by using the back propagation.
The first term in Eq.(\ref{loss}) trains the NN to decrease the time dependence of the dressed Hamiltonian, and the second term in Eq.(\ref{loss}) trains the NN to get the time-periodicity.\footnote{$||\sum_t K'(t)||=0$ leads to the time-periodicity of the kick operator $K(t=0)=K(t=T)$, which further leads to the time-periodicity of $K'(t)$ and $H_r(t)$.} Finally, we do the backpropagation to the NN and update the parameters in the NN. We note that the loss function in Eq.~(\ref{loss}) is zero as the minimum when the RNN outputs the exact RF. We also note, in our framework, there are two hyper-parameters $\gamma$ and $\epsilon$. $\epsilon$ decides how the time-periodicity is emphasized and $\gamma$ must be smaller than one for the convergence of the learning. we set $\epsilon = 20.0, \gamma=0.95$ in this letter. 

In step 1, we numerically derive the kick operator $K(t)$ in the matrix form. However, of course, this numerical form cannot be said to be a ``theoretical analysis method'' because it changes as the parameter or model changes. In step 2, we physicists interpret the results as the operator form. Once we get the kick operator and the RF in the operator form, its construction method does not change if we change the parameters and the model. 

Although it is usually challenging to find the operator form from the matrix form, we can get some hints. 
In the first place, because the width of the input and output layer is limited by the memory of the computers,  we have to use the small system, and it is comparably easy to translate the matrix form to the operator form in small systems.
Moreover, we can get the numerical results in the different parameters, and therefore, we can find the parameter dependence of each matrix element of the kick operator. It tells us what kind of operator constructs each matrix element. We believe these clues are enough for physicists to construct the kick operator in the operator form and find the construction methods.

After deriving the operator form, we can check whether the constructed RF can be applied to other parameters or models. \footnote{For example, after we derive the construction method for the appropriate rotating frame in the two spin model in Eq.(\ref{FM_Kt}), we apply it to the spin-1 model by substituting $H_0$ and $V(t)$ of the spin-1 model, and calculate the dressed Hamiltonian and its eigenvalues. We check whether the time-dependence of the dressed Hamiltonian becomes smaller than the bare Hamiltonian by some order. If not, it might mean that, due to the symmetry of the two-spin model, some parts of the operator form are omitted, and we should derive the construction method in a more complex model with lower symmetry.}

Next, using these procedures, we demonstrate that we ``derive'' the high-frequency expansion method to construct the appropriate RF in a concrete model.

\

{\it{model calculation}} --
Here we introduce the concrete model and demonstrate the concrete procedure. For simplicity, we analyze an interacting two-spin model under the periodic driving, which reads,
\begin{eqnarray}
    \hat{H}_0 &=& \bm{h}\cdot\sum_{i=1,2} \hat{\bm{s}}_i + \sum_{\alpha=x,y,z} J_{\alpha}\hat{s}^{\alpha}_1 \hat{s}^{\alpha}_2,\label{H_0}\\
    \hat{V}(t) &=& \bm{\xi}\cdot \sin(\Omega t) \sum_{i}\hat{\bm{s}}_i,\label{Vt}
\end{eqnarray}
where $\hat{s}_i$ represents the quantum spin operator at site $i$, $\bm{h}$ represents a static magnetic fields, $J_{\alpha}$ represents an interaction between the two spins, and $\bm{\xi}$ represents an AC magnetic fields. For the simplicity, we set $\bm{h}=(0, 0, h_z)$, $\bm{J} = (J_x,0,J_z)$, and $\bm{\xi} = (\xi, 0, 0)$.

This Hamiltonian can be described by a 4*4 matrix in the basis as
\begin{eqnarray}
    H_0 &=& -h_z(\tau^0\otimes\sigma^z + \tau^z\otimes\sigma^0)\nonumber\\
    && - J_x(\tau^x\otimes\sigma^x) - J_z (\tau^z\otimes\sigma^z)\label{H0_mt}\\
    V(t) &=& -\xi \sin(\Omega t) (\tau^0 \otimes \sigma^x + \tau^x\otimes\sigma^0),\label{V_mt}
\end{eqnarray}
where $\tau^\alpha, \sigma^\alpha$ is the Pauli matrix about the site index and the spin index. We vectorize and input this matrix into the NN and perform the learning.

\begin{figure}
    \centering
    \includegraphics[width=0.9\linewidth]{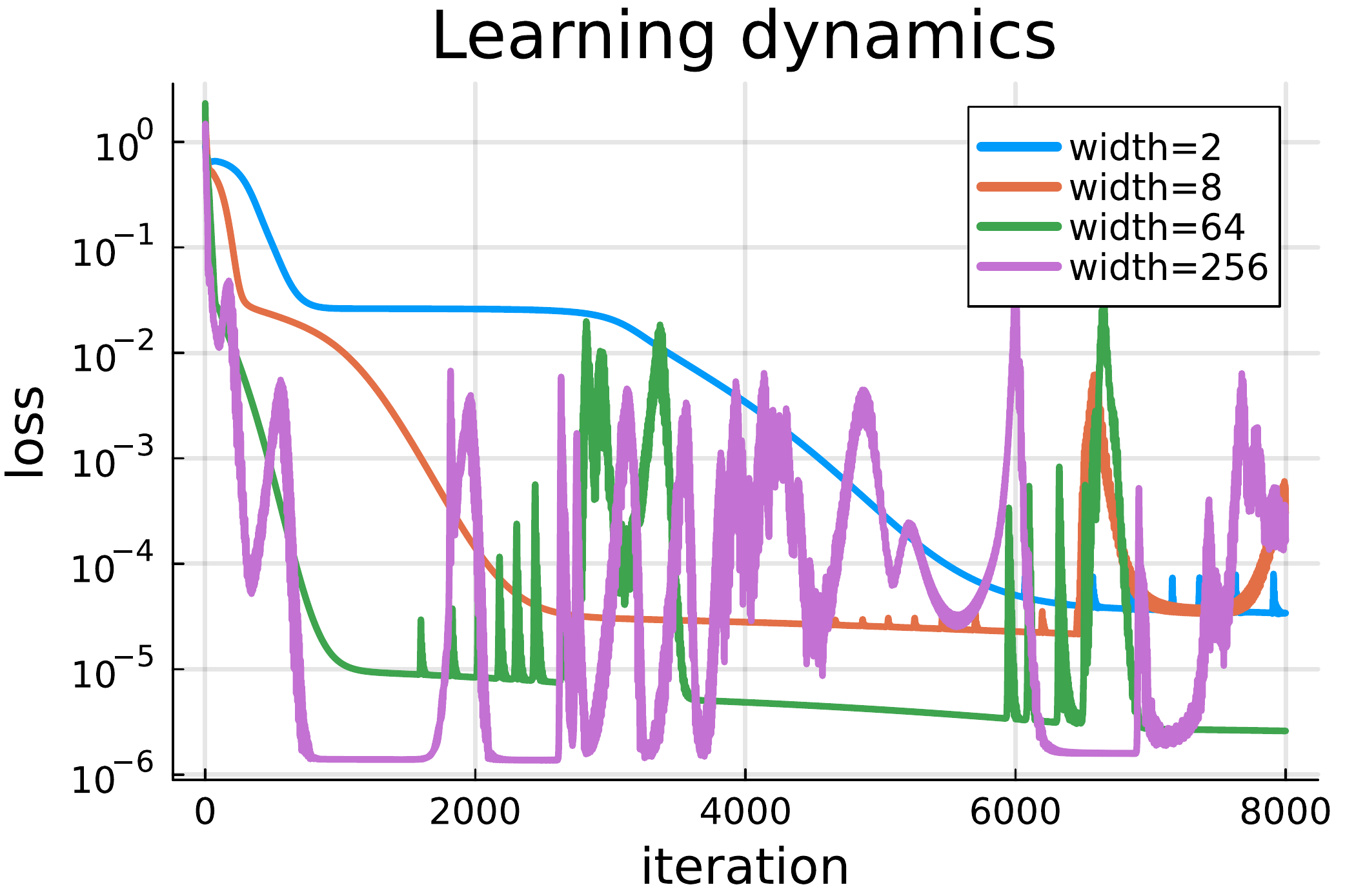}
    \caption{loss function evolution in the learning step and its dependence on the width of the hidden layer.}
    \label{fig:loss_it}
\end{figure}

\begin{figure}
    \centering
    \includegraphics[width=0.98\linewidth]{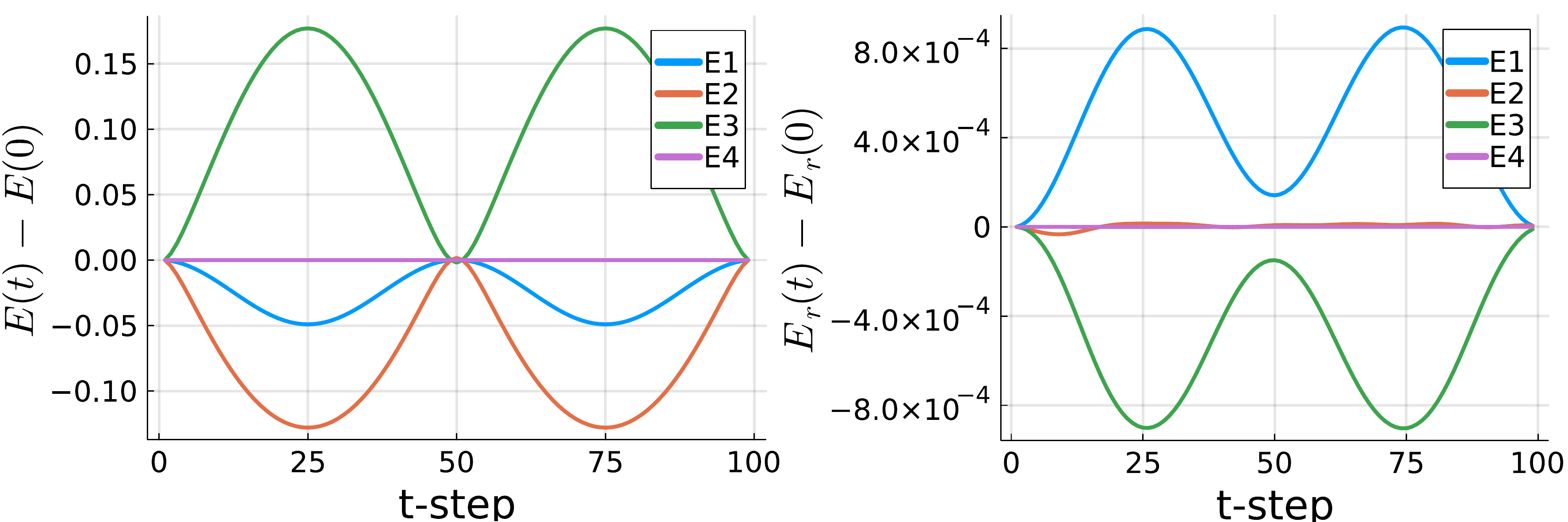}
    \caption{Reduction of the time-dependence of the eigenvalues of the dressed Hamiltonian. The left and right panel show the eigenvalues of $H(t)$ and $H_r(t)$ in the RF derived by the RNN.}
    \label{fig:Hr_reduction}
\end{figure}

Next, we show the concrete results of the two-spin model in step 1. 
We use the NN, in which the size of the input layer and output layer is 20(4 for the parameters of $\hat{H}(t)$ and 16 for  $\hat{K}(t)$) and 16 (16 for  $\hat{K'}(t)$). The NN also holds the two hidden layers. We use Adam \cite{https://doi.org/10.48550/arxiv.1412.6980} as the optimization function and set the iteration number of the learning to 8000, and the activation function is tanh.
Figure~\ref{fig:loss_it} shows the hidden layer width dependence of the learning dynamics. In the single iteration, the NN learns the single cycle of the time-periodic dynamics. In the learning dynamics, the loss function sometimes suddenly increases, and this may stem from the exploding gradient problem in machine learning with an RNN\cite{279181,pmlr-v28-pascanu13}.
We can see that the narrower hidden layers get more stable learning dynamics. We note that we have calculated for some initial parameters of NN and show the best learning dynamics in this letter. We note that the wide hidden layer often makes the learning not converge. (see Supplementary materials (SM) for the details\cite{supply}.)  
The wide hidden layer can also output a too complicated kick operator to use or for us to understand in the operator form. Therefore, the narrow hidden layer might be suitable for our proposed procedure. 
Figure~\ref{fig:Hr_reduction} shows that, in the RF derived by the RNN, the time-dependence of the eigenvalues of the dressed Hamiltonian becomes smaller than that of $H(t)$ by $10^{-2}$ order.

\begin{figure}
    \centering
    \includegraphics[width=0.49\linewidth]{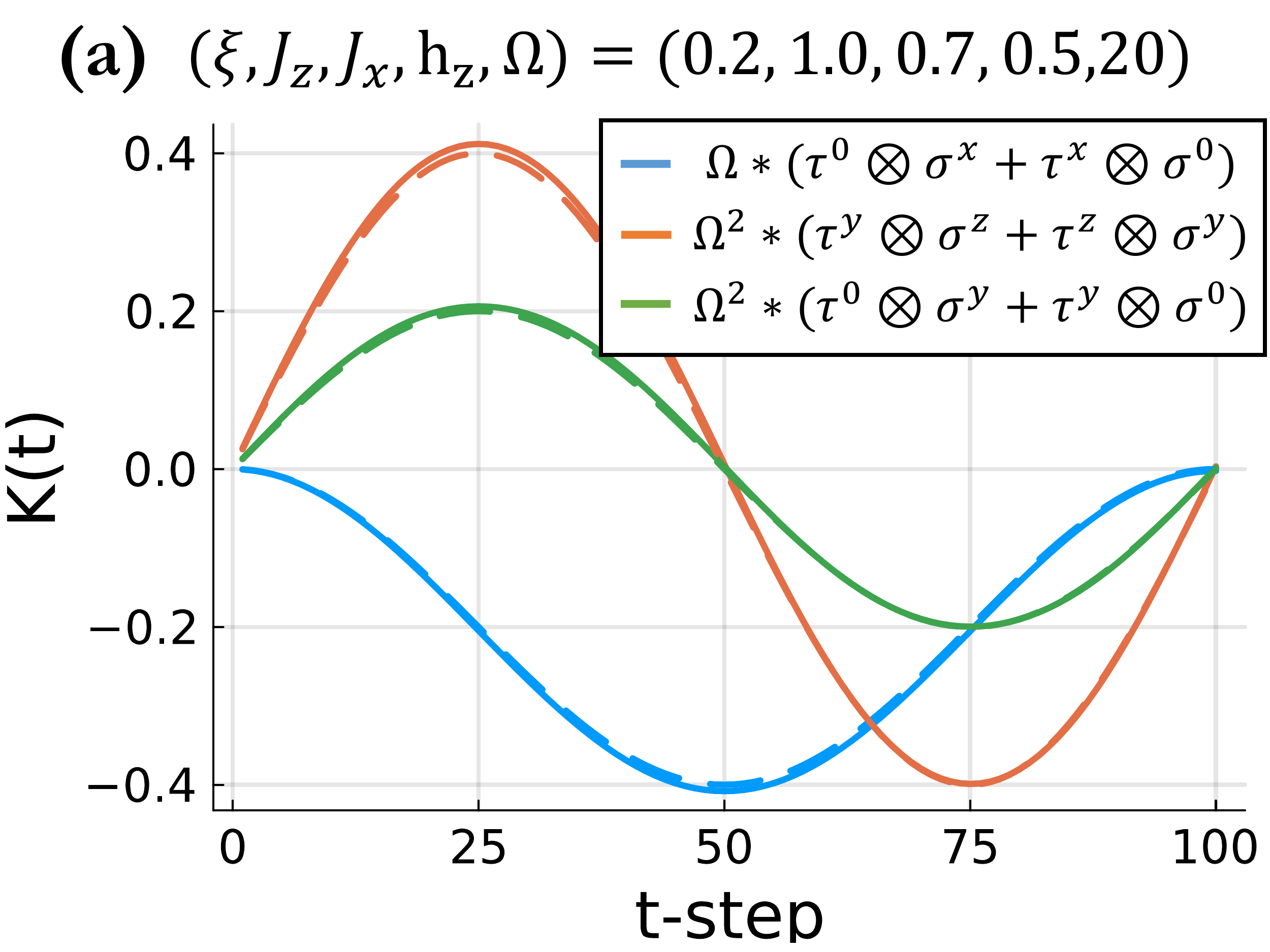}
    \includegraphics[width=0.49\linewidth]{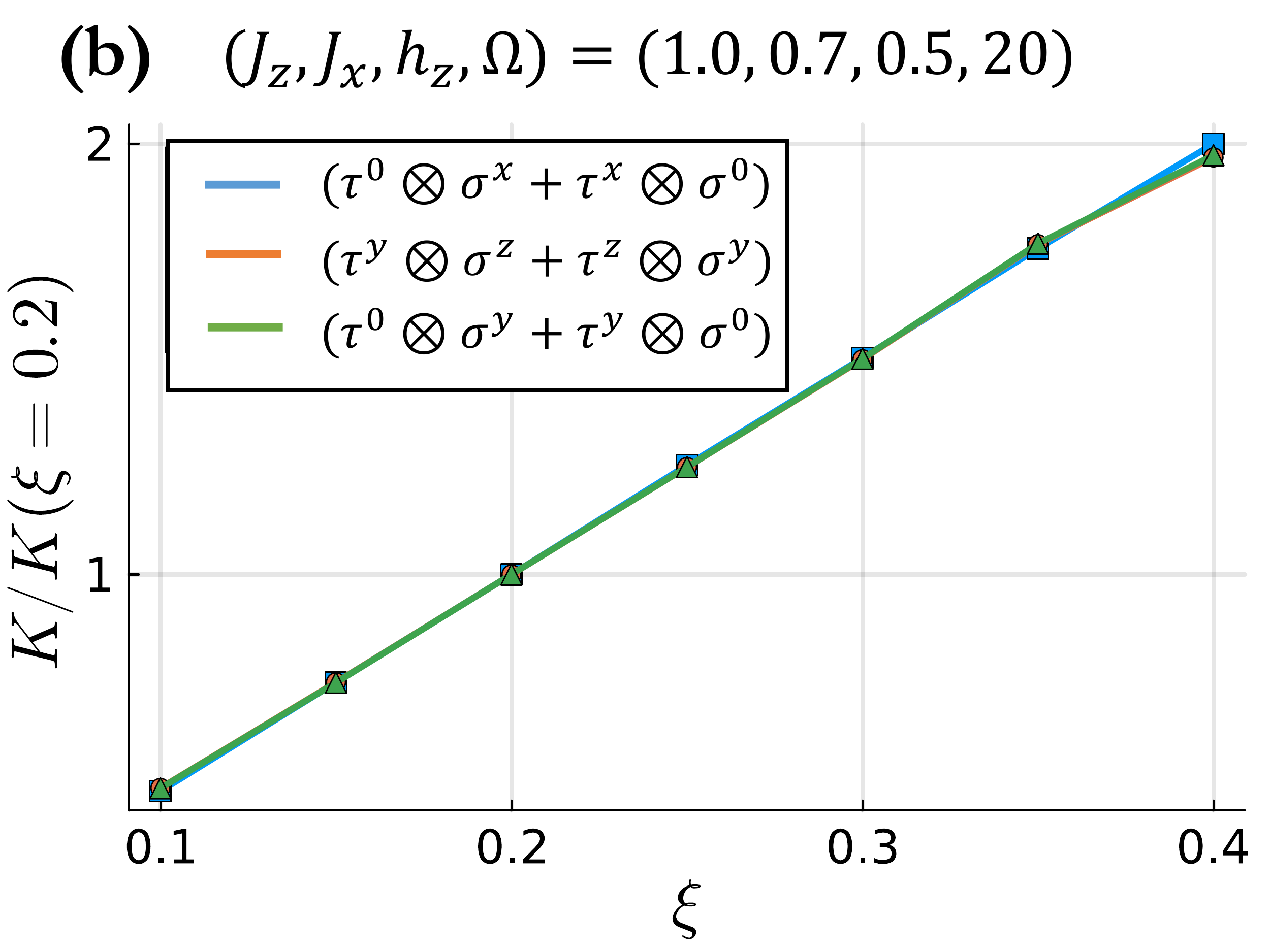}
    \includegraphics[width=0.49\linewidth]{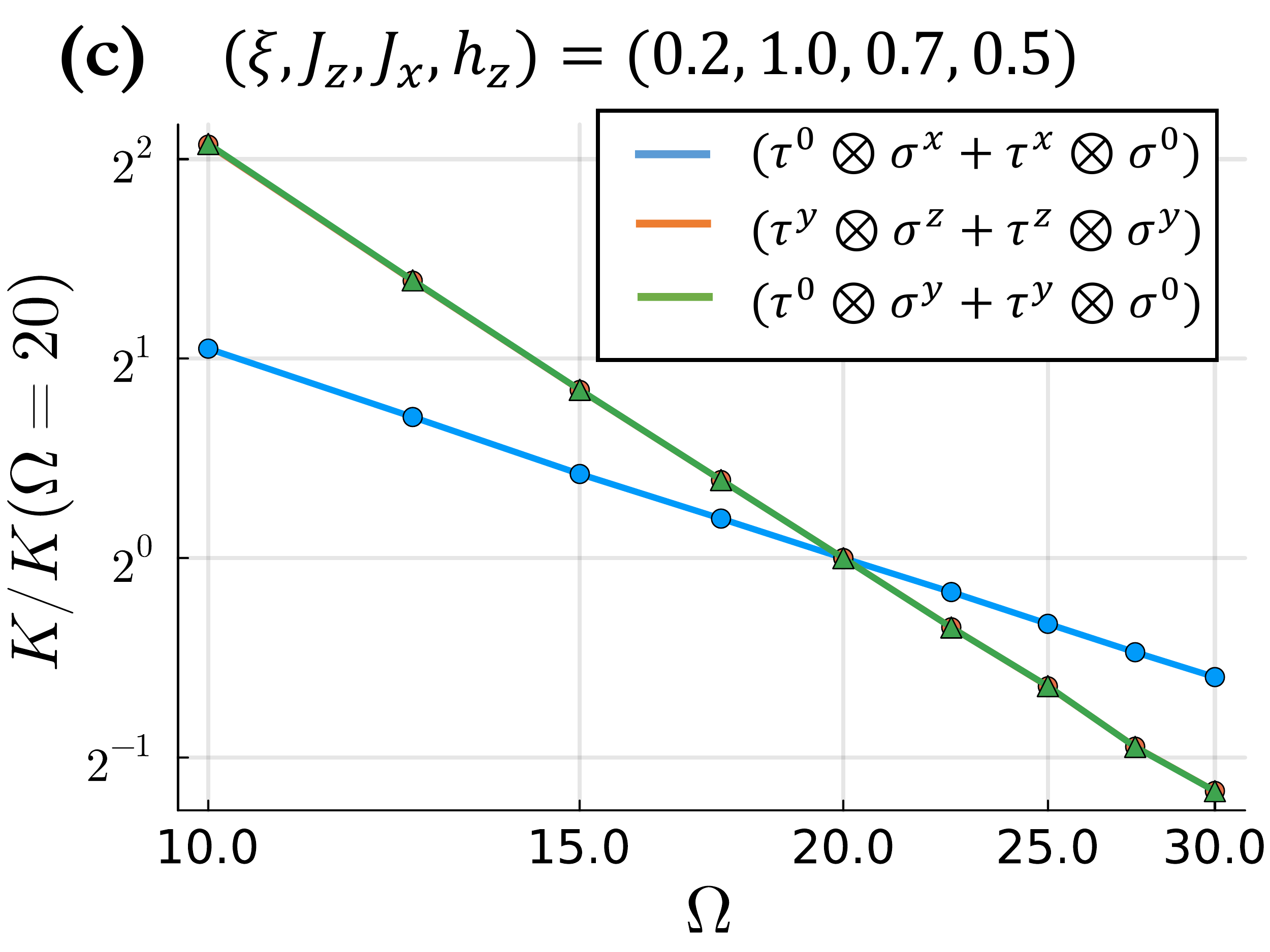}
    \includegraphics[width=0.49\linewidth]{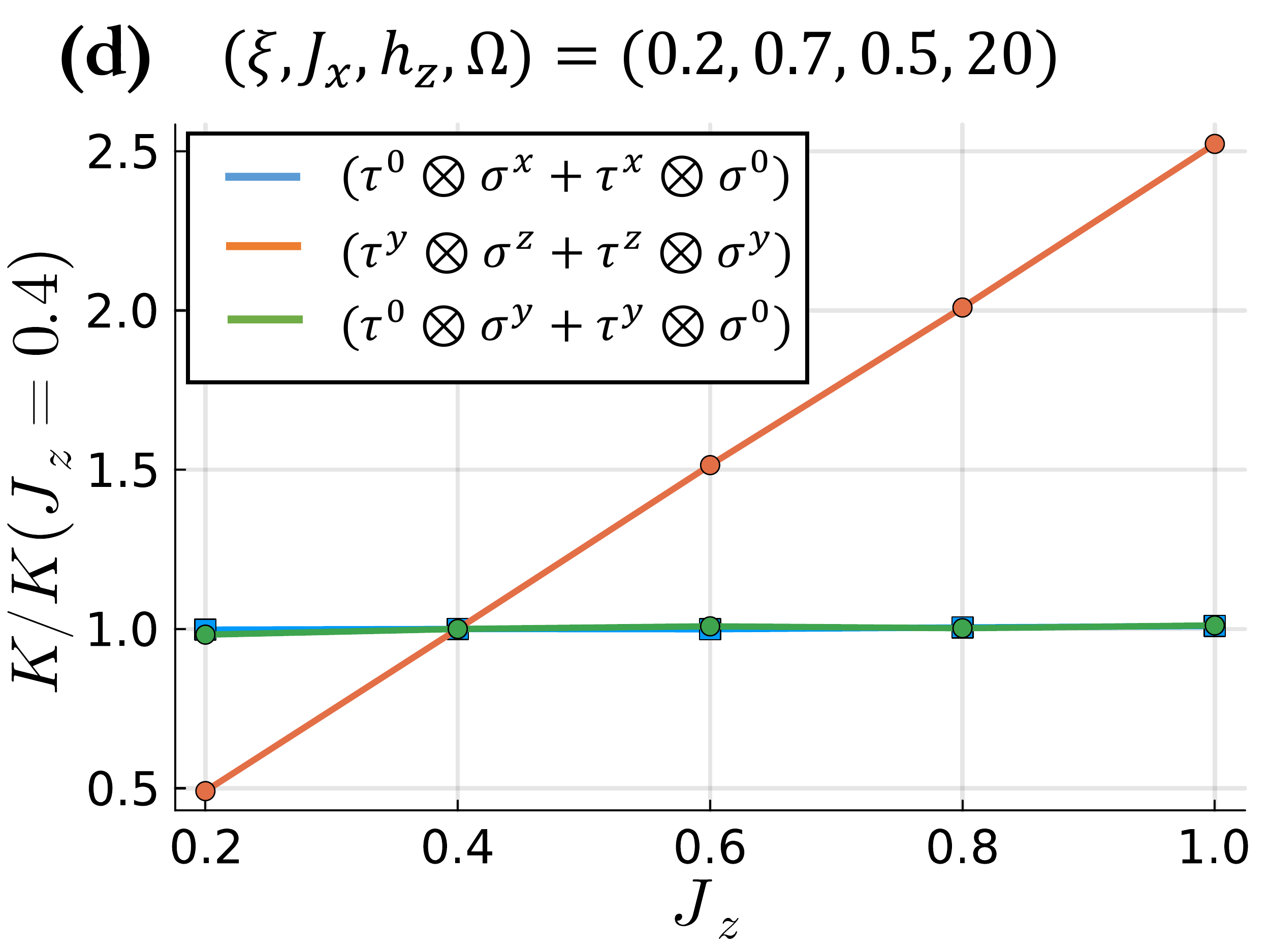}
    \includegraphics[width=0.49\linewidth]{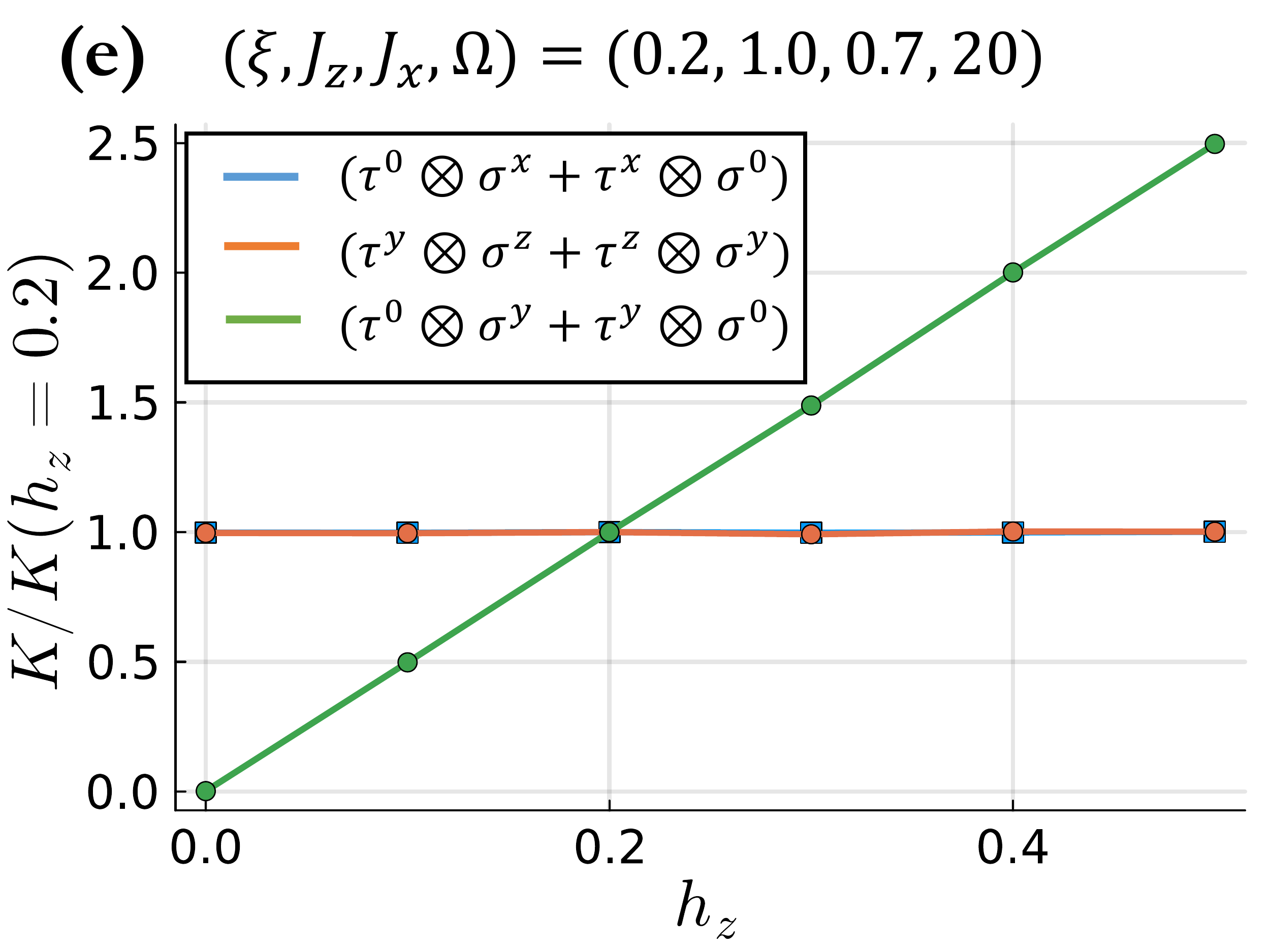}
    \includegraphics[width=0.49\linewidth]{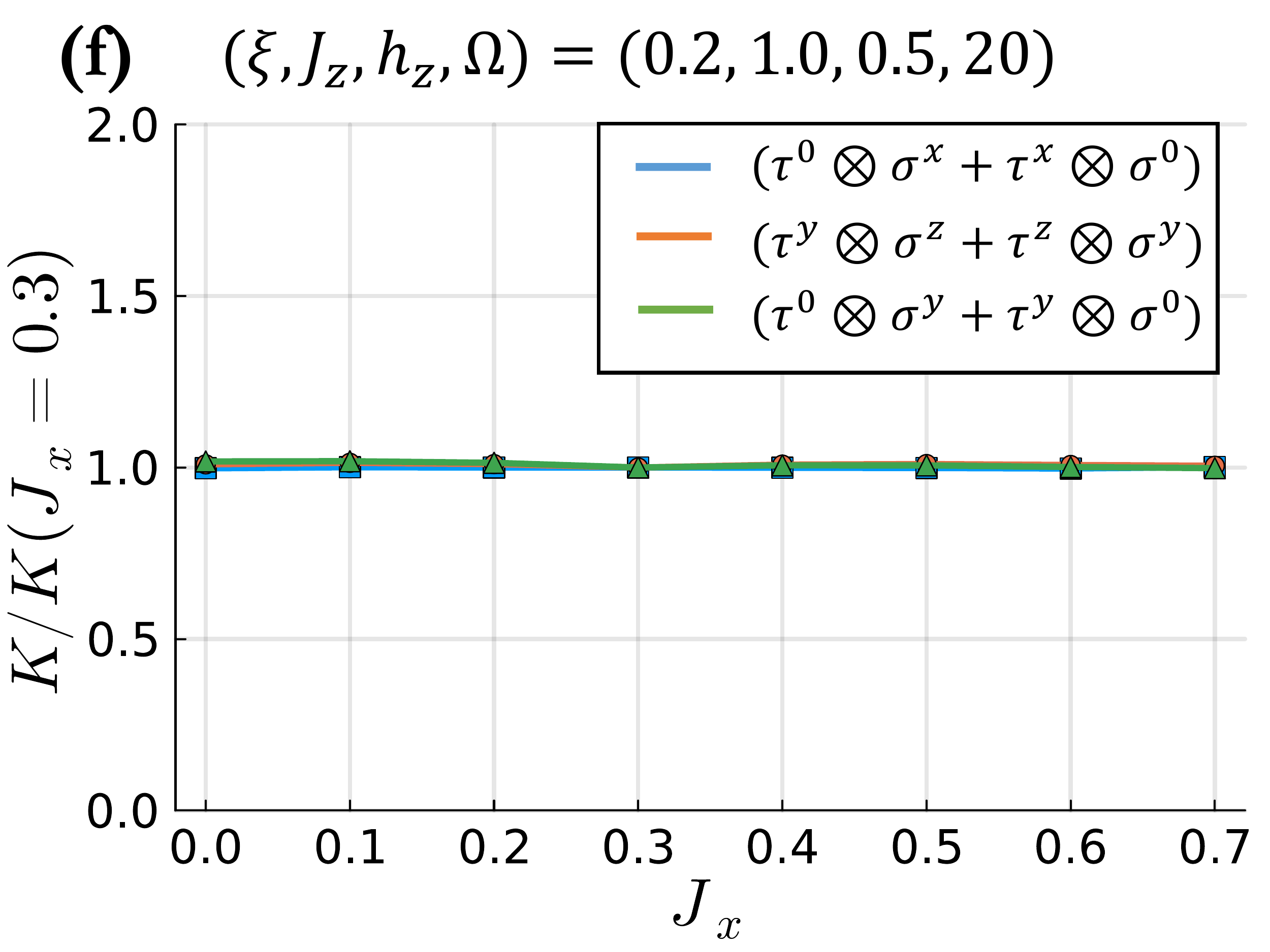}
    \caption{Resulting the kick opeartor and its parameter dependence. Figure~\ref{fig:KT_RF}(a) shows the dominant terms of the resulting kick operator. Figure~\ref{fig:KT_RF}(b)-(f) show the parameter dependence of the amplitude of each dominant term. In order to see the parameter dependence, the amplitude is normalized by the amplitude of certain parameters.}
    \label{fig:KT_RF}
\end{figure}
Then, we show the resulting RF, which is derived by the RNN in figure~\ref{fig:KT_RF}(a). We can find three dominant contributions, which are proportional to (i)$(\tau^0\otimes\sigma^x+\tau^x\otimes\sigma^0)$, (ii)$(\tau^y\otimes\sigma^z+\tau^z\otimes\sigma^y)$, and (iii)$(\tau^0\otimes\sigma^y+\tau^y\otimes\sigma^0)$.
We also check their dependence on the parameters. From figure~\ref{fig:KT_RF}(b)-(f), we can see that each term (i)-(iii) is proportional to $\xi\Omega^{-1}$, $\xi J_z\Omega^{-2}$, and $\xi h_z \Omega^{-2}$. We note that the slight deviation from the proportional relationship should stem from the higher order expansions, and we do not focus on them in this letter although we can analyze them.
Next, we can deduce the operator form of each term. First, let us consider the first term (i). It is independent of the parameters of $\hat{H}_0$ and is proportional to the drive strength, which means it can be described by $\hat{V}(t)$. Moreover, it seems proportional to $(1-\cos(\Omega t))/\Omega$, while $\hat{V}(t)$ is proportional to $\sin(\Omega t)$. Therefore, we can deduce its operator form as $\int dt' \hat{V}(t')$.\footnote{One might think, for example, $\frac{1}{\Omega}(1 - \frac{1}{\Omega}\frac{d}{dt}\hat{V'}(t))$ is also a candidate. However, when we think the driving proportional to the step function about time, this kind of definition is not proper because the time derivative diverges.}  Then, we analyze the second term (ii). Because it is proportional to $\xi J_z$, it should be the product of $\hat{V}(t)$ and $J_z\hat{s}^z_1\hat{s}^z_2$, and actually $(\tau^y\otimes\sigma^z+\tau^z\otimes\sigma^y) = i[(\tau^0\otimes\sigma^x+\tau^x\otimes\sigma^0), \tau^z\otimes\sigma^z]/2$. Moreover, because it is proportional to $\Omega^{-2}\sin(\Omega t)$, it should include the double integration of time. Therefore we can deduce the second term can be written in the operator form as $-i\int dt_1 \int_0^{t_1} dt_2 [\hat{V}(t_2), J_z\hat{s}^z_1\hat{s}^z_2]$.\footnote{One might consider other candidate such as $-2i J_z\int dt_1 \int_0^{t_1} dt_2 \hat{V}(t_2)\hat{s}^z_1\hat{s}^z_2$. However, if it is, $-2iJ_x\int dt_1 \int_0^{t_1} dt_2 \hat{V}(t_2) \hat{s}^x_1\hat{s}^x_2$ term also should appear in the kick operator while in fact there is no such a term. $J_x$-independence of the dominant terms implies that the kick operator should be written in the commutator form. }
By performing the same procedure with the other term, we can reach the proposal construction of the kick operator as
\begin{eqnarray}
    \hat{K}(t) = \int dt_1 \Bigl(\hat{V}(t_1) -i \int dt_2 \bigl[\hat{V}(t_2), \hat{H}_0\bigr]\Bigr).\label{FM_Kt}
\end{eqnarray}
Finally, we can check whether this construction method can be applied to the other models and conclude that this is the appropriate RF. The dashed line in Figure \ref{fig:KT_RF}(a) describes the kick operator in Eq.~(\ref{FM_Kt}), which highly matches the RNN-derived kick operator.(see the SM for the evaluation of deducing the appropriate frame.)
In fact, this derived method corresponds to the Floquet-Magnus expansion up to the second order.
Interestingly, this RNN treatment gives us the appropriate RF, not the exact RF, in which $H_r(t)$ is time-independent, even though the loss function defined in Eq.~(\ref{loss}) becomes the minimum at the exact RF. 
It should stem from the representation capability limit of the NN because the exact RF is usually a highly complex function of the operators. However, this limit of NN is beneficial for physicists because the limited NN should prefer the perturbative or simple RF, and it should be easier for us to interpret and more useful. Therefore, the narrow hidden layers should be better, while they should have enough width to express the perturbative expression of RF.
We note that we also tried to derive the appropriate RF in the resonant regime where the energy scale of the system is comparable to the frequency of the driving.
Approaching to the resonant regime, the learning dynamics become unstable and the RNN rarely achieve the appropriate RF, and we cannot get the appropriate RF at the resonant regime. (see the detail in the SM\cite{supply}.) These results should imply that, in the resonant regime, there is no simple or perturbative construction methods of the appropriate RF.

\

{\it{Generalization of the procedure to derive the appropriate unitary transformation.}} --
Finally, we argue how the procedure we demonstrate in the previous section can be applied to other general problems. In the previous section, we have defined the loss function to train the NN to give us the RF in which the time dependence of the dressed Hamiltonian is small. This idea can be applied to find other appropriate frames or projections, which make us able to solve the physical problem easily. We can define the loss function as it becomes small in the frame which has desirable properties, and then, after the learning (if the learning works well), NN gives us the desirable frame. As we have demonstrated, we first calculate the appropriate frame or the projection in the matrix form in a small system, analyze the parameter dependence, and then, we physicists should be able to translate the operator form and develop the construction method. If the translation is difficult, we can also utilize another reinforcement learning techniques for finding the construction method of the appropriate frame, such as the Monte Carlo tree search algorithm.\cite{10.1007/978-3-540-75538-8_7, DBLP:books/daglib/0016921, pmlr-v70-kusner17a, Silver2017, Lample2020Deep, NEURIPS2021_d073bb8d, sun2023symbolic} In this method, we update the agent's policy to make the final operator form almost the same as the derived appropriate frame and get the construction method described by the expression tree. Although this method might be able to give us the operator form of the appropriate frame automatically, we have to define the search area of the operation previously. \footnote{For the case of the rotating frame in periodically-driven systems, the commutation of the static Hamiltonian and the driving term, the time integration, and the summation of terms are necessary as possible operations for the above method to derive the high-frequency expansion.} However, in general cases, we usually do not know what operations are necessary to construct the appropriate frame. In such cases, we first deduce the necessary operation from the parameter dependence of the appropriate frame numerically derived at step 1 and apply the above methods to derive how to combine the operation to get the appropriate frame. Therefore, the deduction by the physicist should be necessary independent of whether we use the reinforcement learning techniques.

\

{\it{Conclusion.}} --
In this letter, we have proposed the method to construct the appropriate RF with the aid of RNN and demonstrate the concrete process and the results. 
By defining the loss function as it becomes small when the output RF has the desirable property, which is the smallness of the time dependence of the dressed Hamiltonian, we can get the appropriate RF in the matrix form. In a small system, we can translate the results to the operator form and get the analytical method to construct the appropriate RF.
Interestingly, the proposed method gives us the appropriate RF, not the exact RF. This property should stem from the representation capability limit of the NN, and this limit is beneficial for physicists to interpret and utilize the results. Although some previous works\cite{PhysRevX.10.011006, PhysRevResearch.2.023358, doi:10.1021/acs.jctc.2c00702,https://doi.org/10.48550/arxiv.2206.12363} solve the dynamics or equilibrium state using RNN, our subject and goal differ from theirs. While they focus on solving and getting the numerical results, we aim to derive the theoretical analysis methods. We believe our proposed method is useful even in other systems and this work paves a new way to utilize machine learning to develop analytical methods in physics. 

\

{\it{Acknowledgements.}}--
YM deeply appriciate Takashi Mori and Kaoru Mizuta for fruitiful discussion and Ken Mochizuki and Shuntaro Sumita for their valuable comments on the manuscript. YM also thanks Irasutoya\footnote{ Irasutoya is the japanese free illustration site at https://www.irasutoya.com/} for allowing me to use the illustration of the physicist in Figure~\ref{fig:notionRF}. We use Julia\cite{bezanson2017julia} and its library Flux.jl\cite{Innes2018} for all calculation, and the codes we have used are uploaded in the Github\cite{github}. This work is supported by RIKEN Special Postdoctoral Researcher Program.

\bibliographystyle{apsrev4-1}
\bibliography{ML.bib}
\clearpage

\renewcommand{\thesection}{S\arabic{section}}
\renewcommand{\theequation}{S\arabic{equation}}
\renewcommand{\thefigure}{S\arabic{figure}}

\setcounter{equation}{0}
\setcounter{figure}{0}

\onecolumngrid
\begin{center}
{\large
{\bfseries Supplemental Materials for \\ ``Machine-learning-assisted construction of appropriate rotating frame'' }}
\end{center}

\vspace{10pt}

\onecolumngrid

\section{S1. \ \ \label{app:Wdep} hidden-layer-width dependence of the learning dynamics}
\begin{figure}
    \centering
    \includegraphics[width=0.48\linewidth]{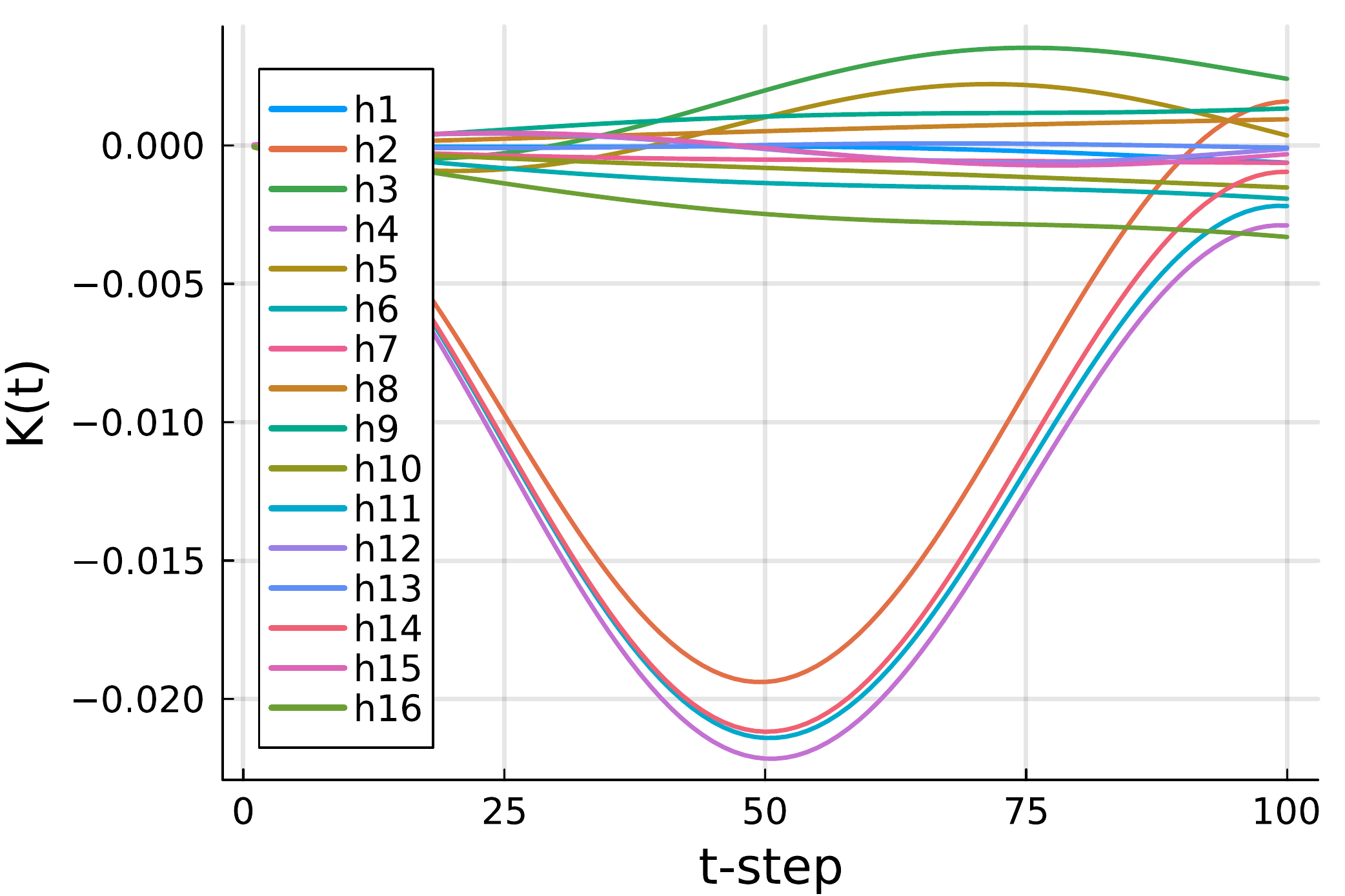}
    \includegraphics[width=0.48\linewidth]{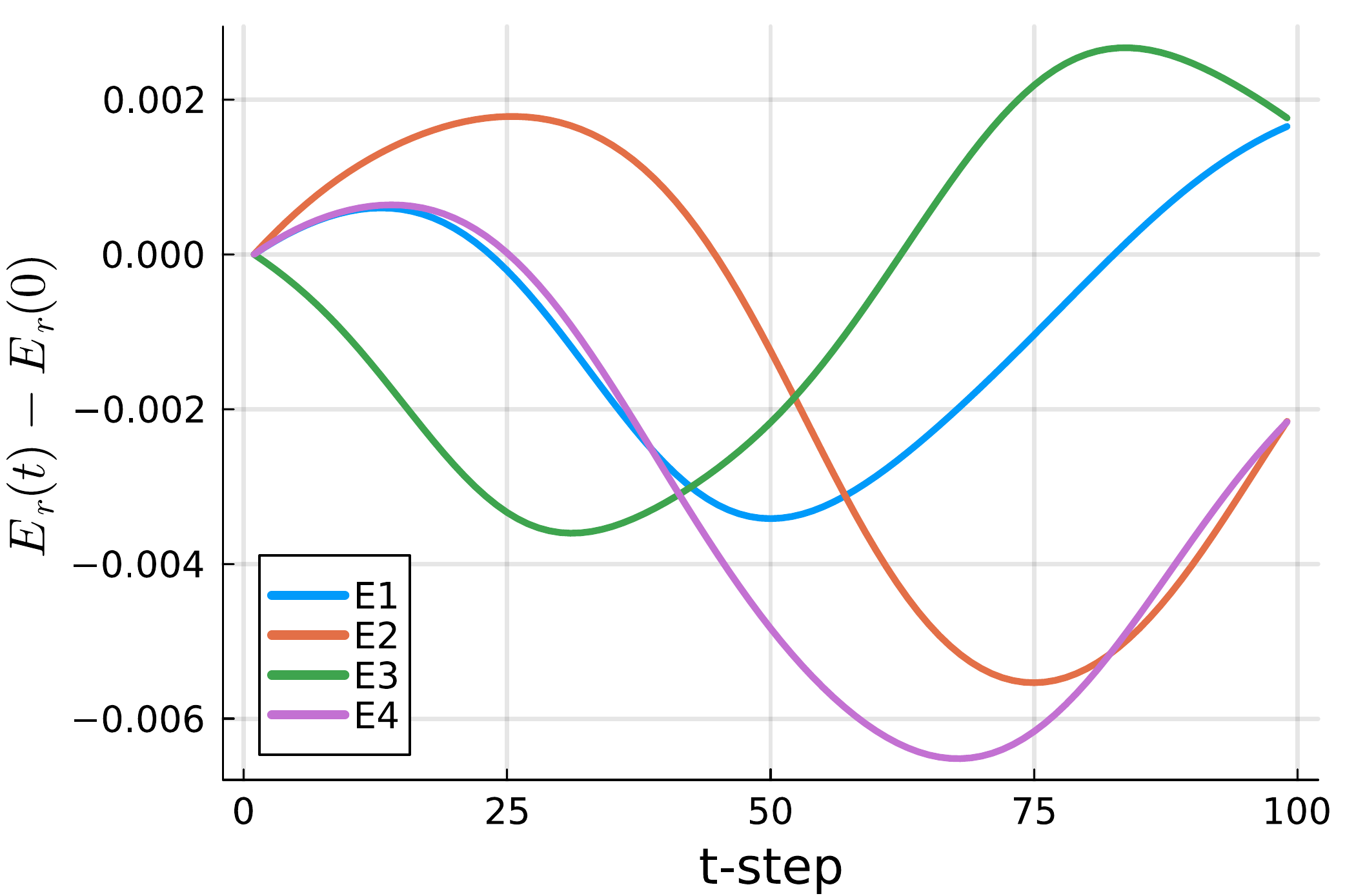}
    \caption{The results in $w=64, b=1$, where the learning ends when $l\sim 10^{-2}$. In such case, the resulting kick operator has many nonzero elements, and the resulting dressed Hamiltonian does not get the time periodicity.}
    \label{fig:bad_ex}
\end{figure}

In figure.~\ref{fig:loss_it} in the main text, we have shown only successful learning dynamics. In this section, we show the learning dynamics of 5 batches, in which the NN is randomly initialized and the width of the hidden layers is each 2, 4, 8, 16, 64, and 256. Within these five batches, we can see that the learning dynamics are most stable when the width of the hidden layer is 8. As we have noted in the main text, narrow hidden layers are proper for the stable learning dynamics while it rarely get the appropriate RF when it is too narrow.
In figure.~\ref{fig:bad_ex}, we also show the resulting kick operator and the dressed Hamiltonian when the learning does not work well. We note that we vectorize a 4*4 Hermitian matrix as
\begin{eqnarray}
    M = \begin{pmatrix}
        \mr{h1} & \mr{h2}+i\mr{h3} & \mr{h4}+i\mr{h5} & \mr{h6}+i\mr{h7} \\
        \mr{h2}-i\mr{h3} & \mr{h8} & \mr{h9}+i\mr{h10} & \mr{h11}+i\mr{h12} \\
        \mr{h4}-i\mr{h5} & \mr{h9}-i\mr{h10} &\mr{h13} & \mr{h14}+i\mr{h15}\\
        \mr{h6}-i\mr{h7} & \mr{h11}-i\mr{h12} & \mr{h14}-i\mr{h15} & \mr{h16}
    \end{pmatrix}.\nonumber\\
\end{eqnarray}
In figure~\ref{fig:bad_ex}, the time-dependence of each element is shown.

\begin{figure}[b]
    \centering
    \includegraphics[width=0.46\linewidth]{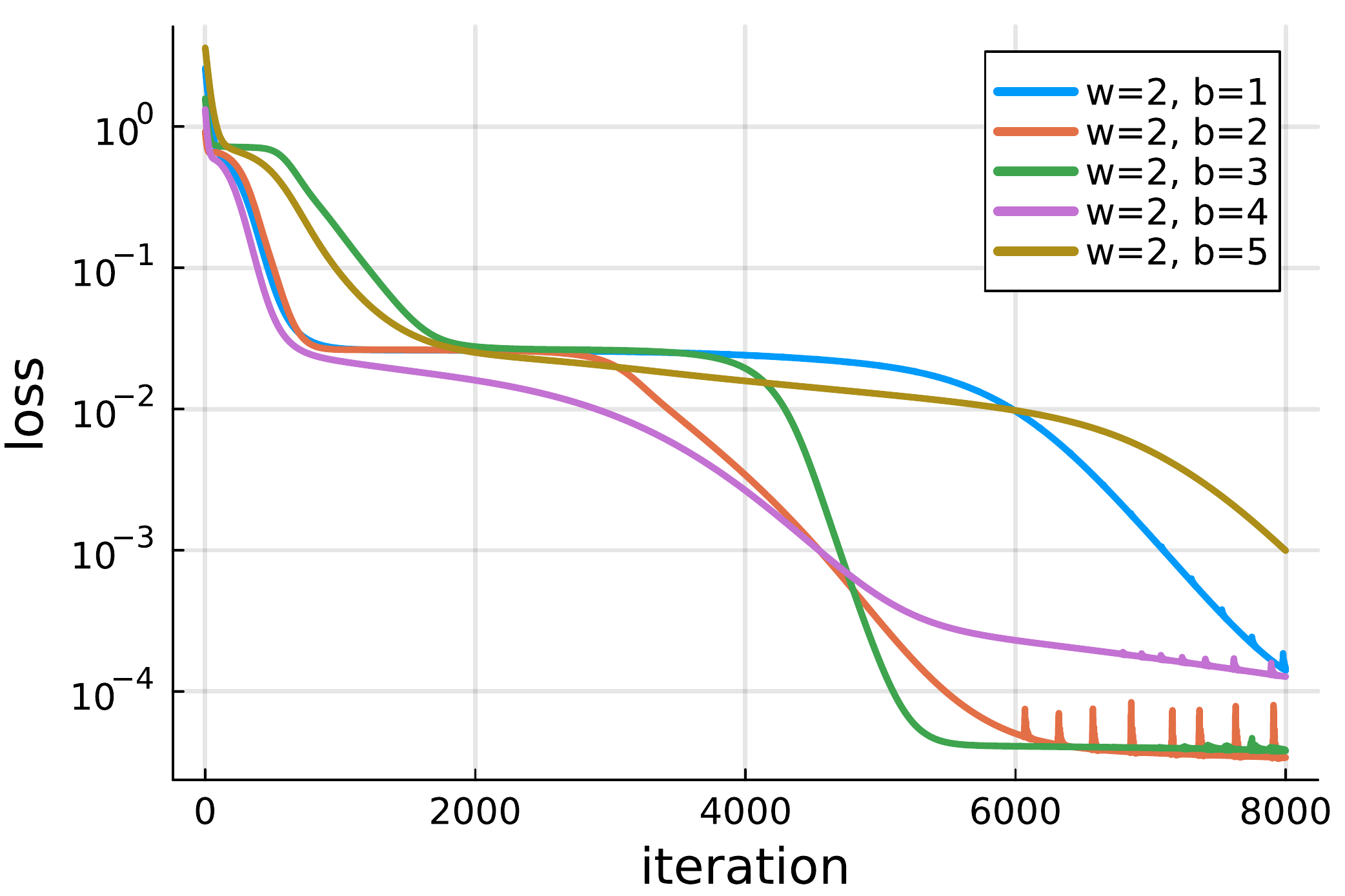}
    \includegraphics[width=0.46\linewidth]{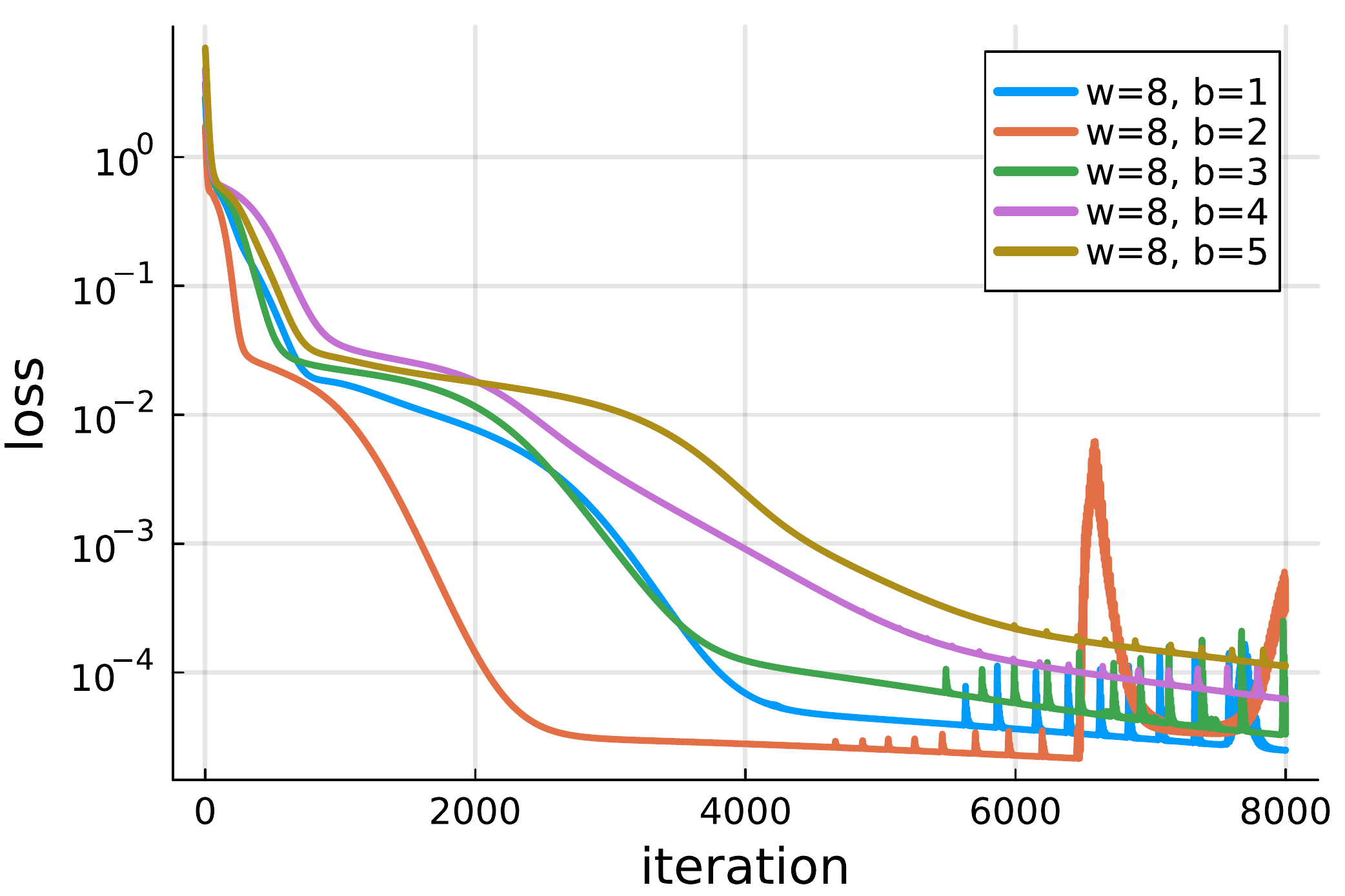}
    \includegraphics[width=0.46\linewidth]{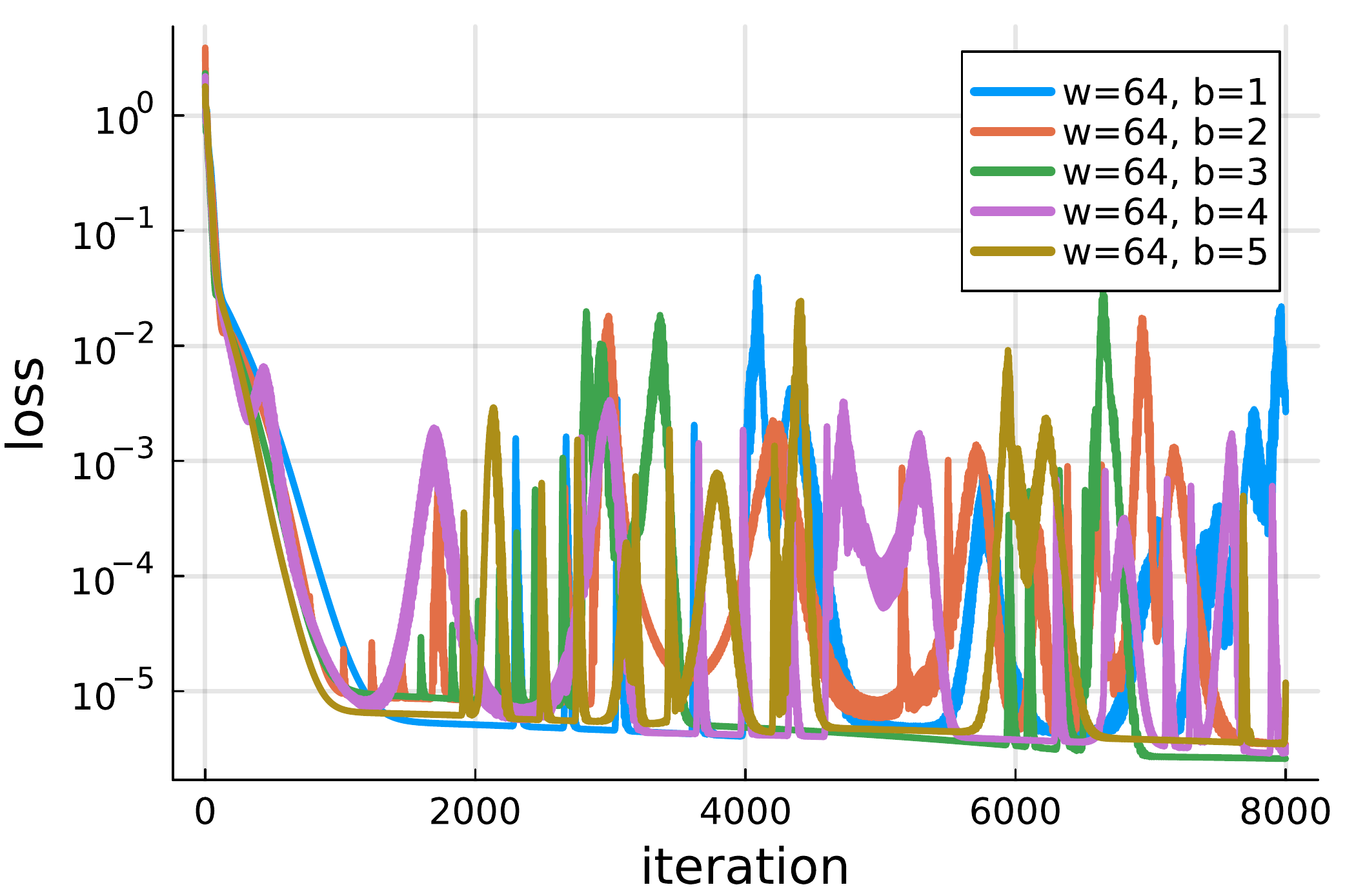}
    \includegraphics[width=0.46\linewidth]{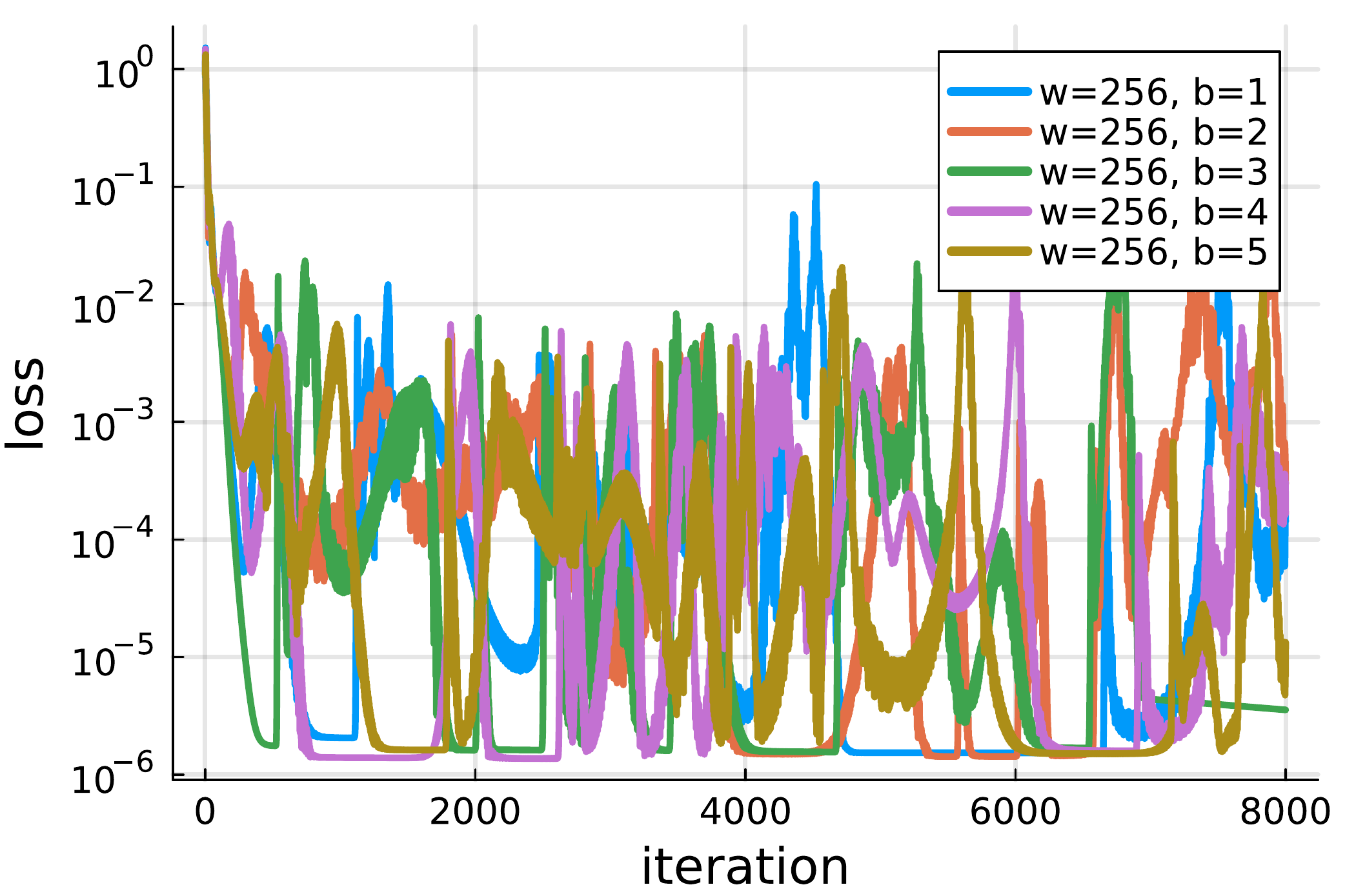}
    \caption{The hidden layer width dependence of the learning dynamics. We represent $w$ as the width of the hidden layer, and $b$ as the batch index. The parameters are $(\xi, \Omega, J_z, J_x, h_z) = (0.2, 20.0, 1.0, 0.7, 0.5)$.}
    \label{fig:batchdep}
\end{figure}

\section{S2. \ \ \label{app:eval_acc} Evaluation of deducing the operator form}
\begin{figure}
    \centering
    \includegraphics[width=0.98\linewidth]{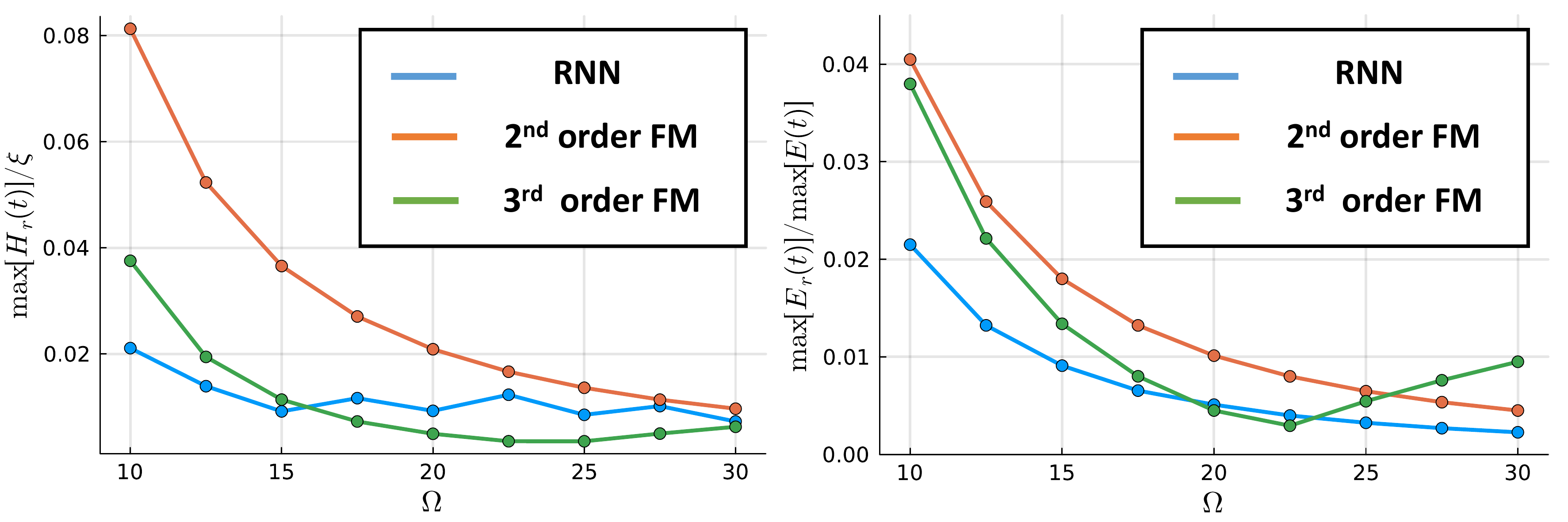}
    \caption{check the derived construction method reproduce the numerical results by the RNN.\\
    The left figure shows that the max value of the dressed Hamiltonian, and the right shows that the max value of the eigenvalues of the dressed Hamiltonian. Both figure shows that the second order perturbation with the Floquet-Magnus expansion realize $10^{-2}$ order compared to the bare Hamiltonian.The parameters are $(\xi, J_z, J_x, h_z) = (0.2, 1.0, 0.7, 0.5)$.}
    \label{fig:amp_check}
\end{figure}

In this section, we discuss how to evaluate the deduce. Because we use the NN as a function approximator, the important thing is not to derive the operator form exactly reproducing the numerical results. We want the operator form, which roughly reproduces the numerical results(i), can be written in a simple form(ii), and has the desirable property(iii). We have already shown that the ``derived'' operator form in Eq.(\ref{FM_Kt}) satisfies conditions (i) and (ii). Thus we check condition (iii), which is the small time-dependence of the dressed Hamiltonian. Figure \ref{fig:amp_check} shows the maximum value of the dressed Hamiltonian divided by that of the bare Hamiltonian and the maximum value of the eigenvalues divided by that of the bare Hamiltonian. Both figures show that the "derived" operator form can realize a smaller time-dependence than the bare Hamiltonian by $10^-2$ order. Therefore, we can conclude that the derived construction of the rotating frame is appropriate enough. This evaluation method of the derivation of the operator form is also applicable to general cases. Figure \ref{fig:amp_check} also shows that the RNN-derived rotating frame is comparable to the third-order Floquet-Magnus expansions and better than it in the small frequency region $\Omega\sim10$. We note that we can also``derive'' the Floquet-Magnus expansion up to the third order from the RNN-derived kick operator, and discuss it in the next section.

\section{S3. \ \ \label{app:3rd} Derivation of the third-order Floquet-Magnus expansion}
\begin{figure}
    \centering
    \includegraphics[width=0.5\linewidth]{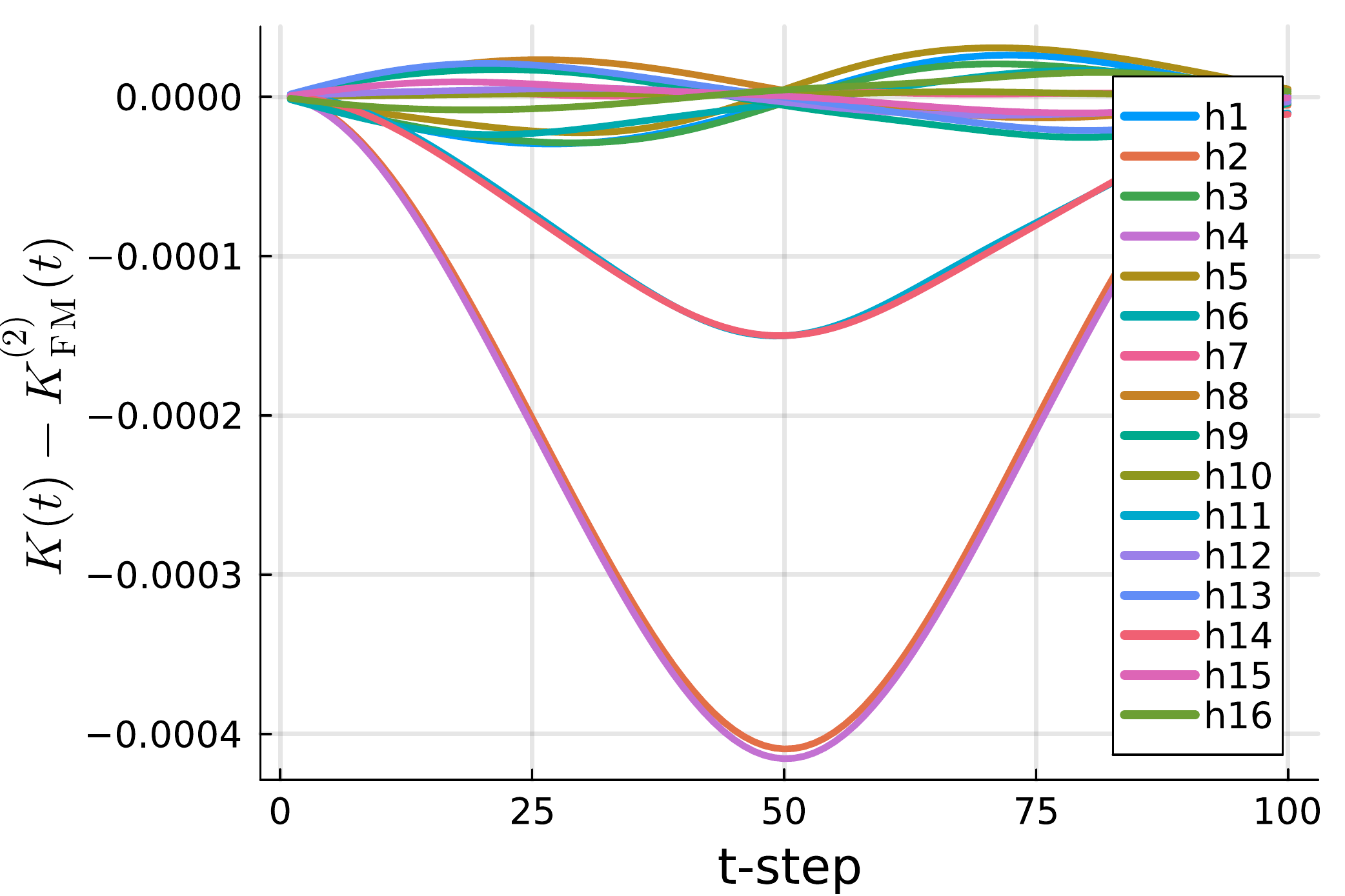}
    \caption{the difference between the kick operator derived by the RNN and by Eq.(\ref{FM_Kt}).}
    \label{fig:KT_diff}
\end{figure}

\begin{figure}
    \centering
    \includegraphics[width=0.32\linewidth]{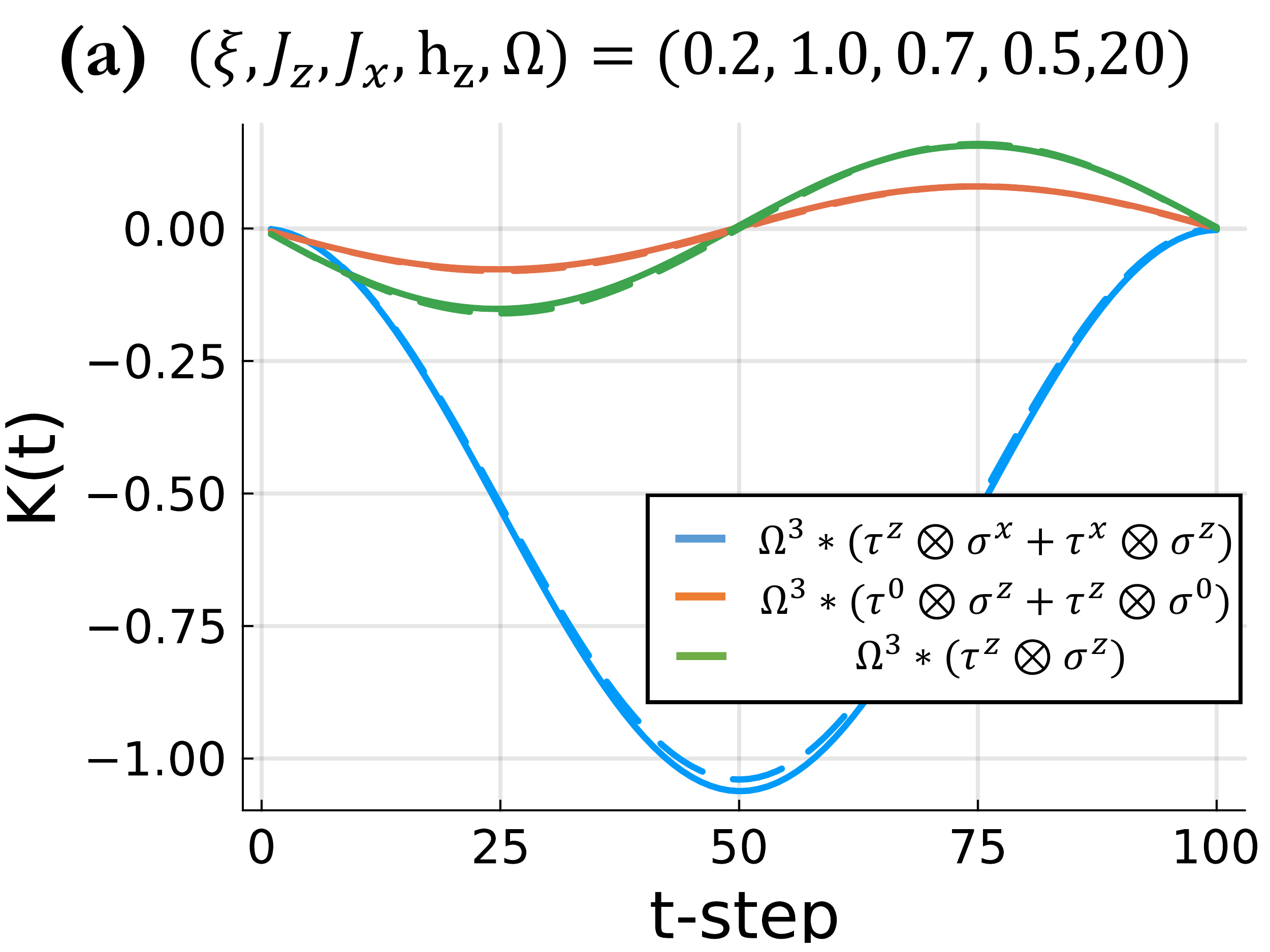}
    \includegraphics[width=0.32\linewidth]{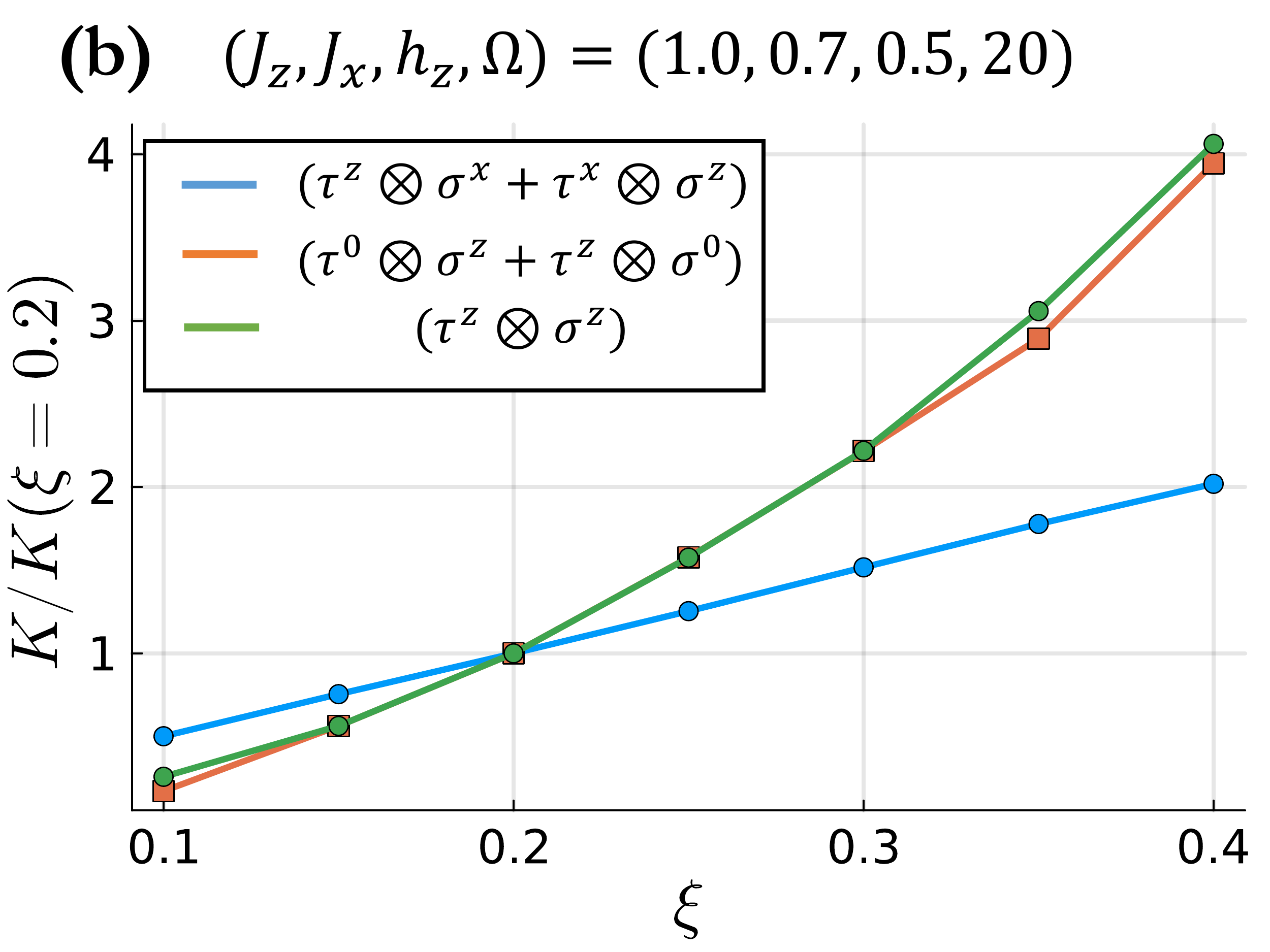}
    \includegraphics[width=0.32\linewidth]{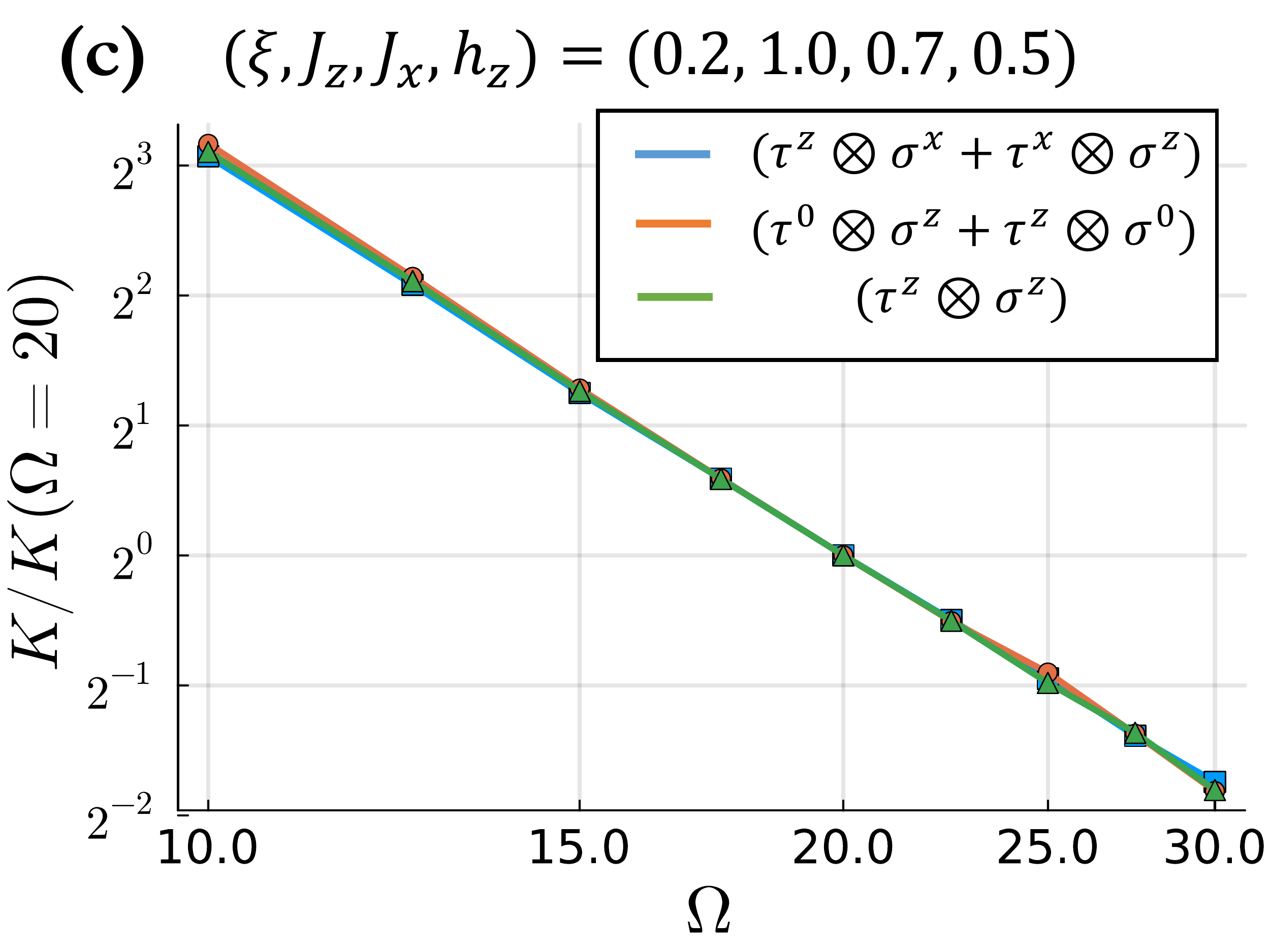}
    \includegraphics[width=0.32\linewidth]{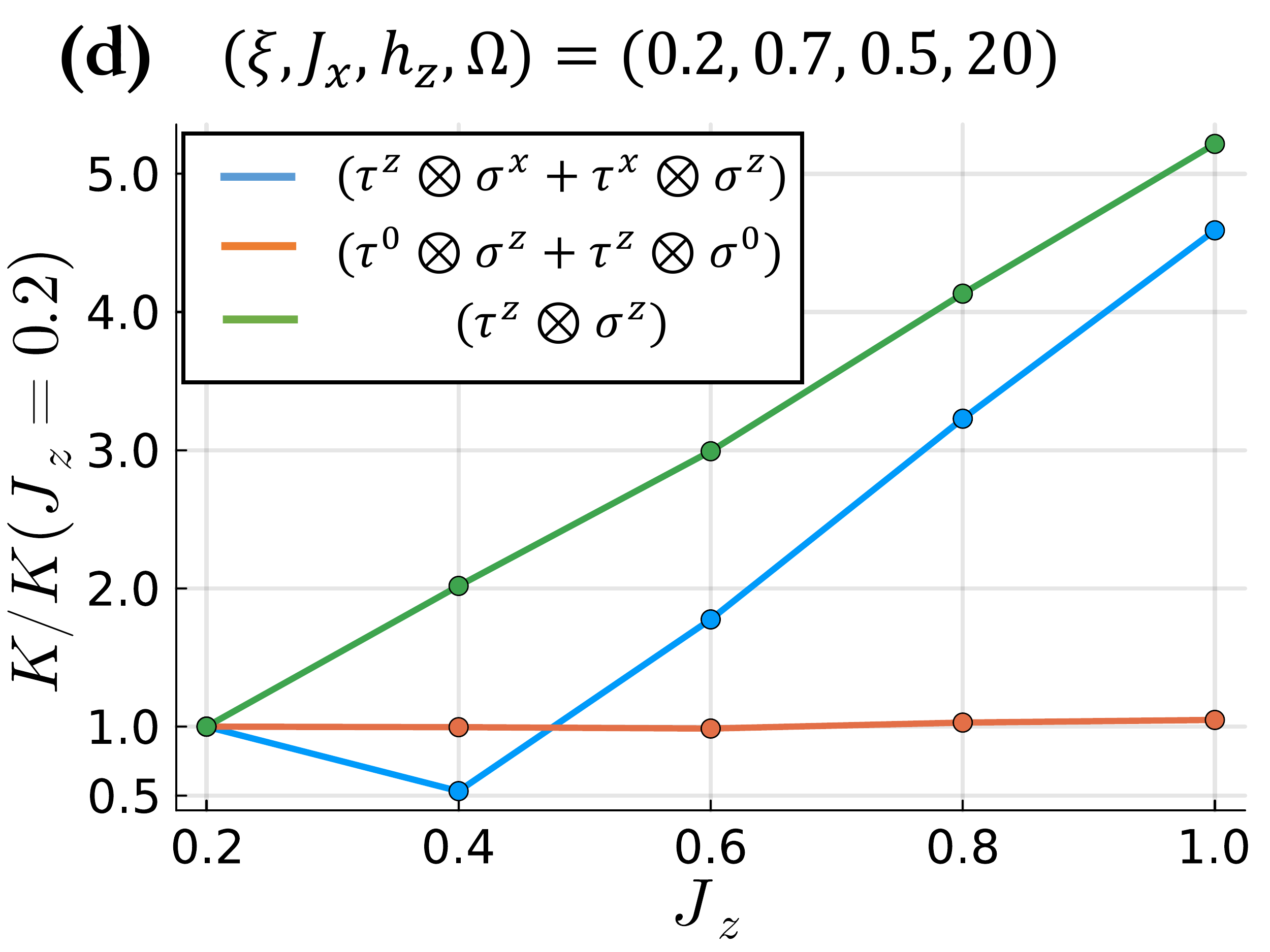}
    \includegraphics[width=0.32\linewidth]{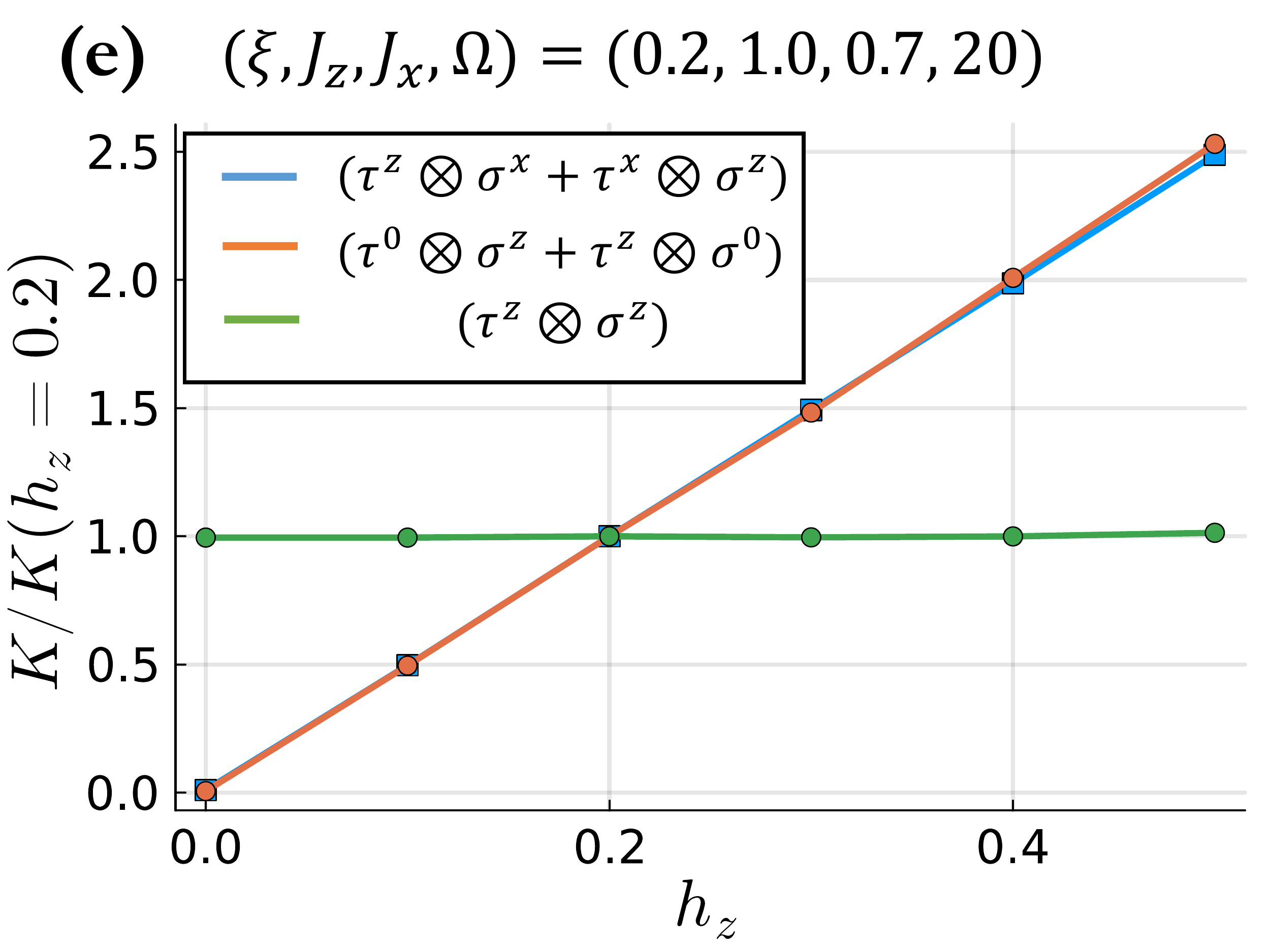}
    \includegraphics[width=0.32\linewidth]{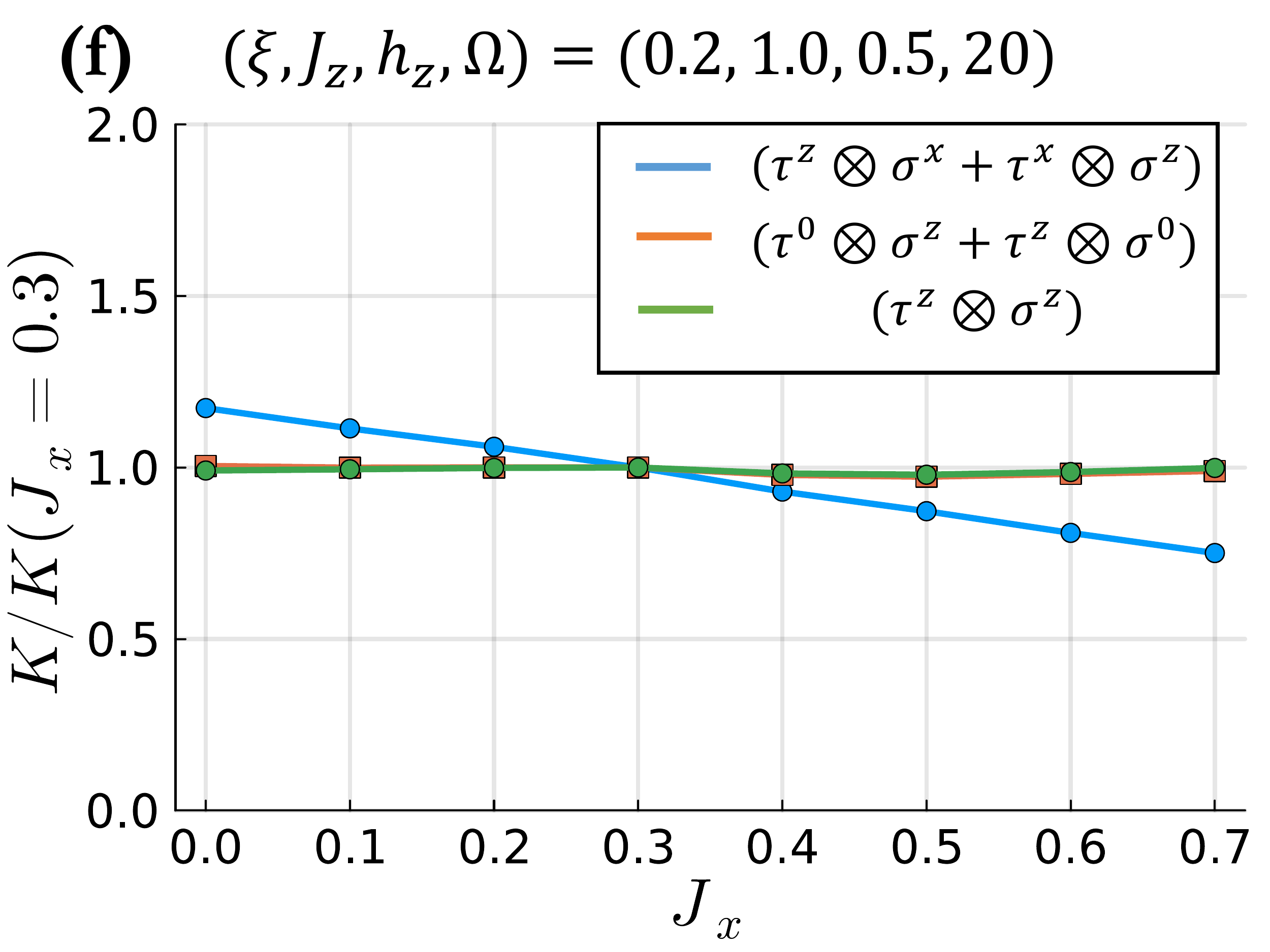}
    \caption{Resulting the kick opeartor and its parameter dependence. Figure~\ref{fig:KT_RF}(a) shows the sub-dominant terms of the resulting kick operator. Figure~\ref{fig:KT_RF}(b)-(f) show the parameter dependence of the amplitude of each dominant term. In order to see the parameter dependence, the amplitude is normalized by the amplitude of certain parameters.}
    \label{fig:KT_RF3}
\end{figure}

When we want a more complex but less time-dependent rotating frame construction method, we focus on the sub-dominant terms in the RNN-derived results. Figure \ref{fig:KT_diff} shows the difference between the RNN-derived Kick operator and the second-order Floquet-Magnus expansion in Eq.(\ref{FM_Kt}). Because we already have reproduced the dominant terms up to the $10^{-3}$ order by the second-order Floquet-Magnus expansion, the rest sub-dominant terms are $10^{-4}$ order. 
Next, we summarize these sub-dominant terms to the three matrix forms as $(\tau^z\otimes\sigma^x+\tau^x\otimes\sigma^z)$,$(\tau^z\otimes\sigma^0+\tau^0\otimes\sigma^z)$ and $(\tau^z\otimes\sigma^z)$, and check the parameter dependence of them. Figure \ref{fig:KT_RF3}(a) shows the time-dependence of the dominant and sub-dominant terms in the kick operator, and Figure \ref{fig:KT_RF3}(b-f) shows the parameter dependence of their amplitude. We can see that the sub-dominant terms ($(\tau^z\otimes\sigma^x+\tau^x\otimes\sigma^z)$,$(\tau^z\otimes\sigma^0+\tau^0\otimes\sigma^z)$ and $(\tau^z\otimes\sigma^z)$) are proportional to $\xi h_z(2J_z-J_x) \Omega^{-3}$, $\xi^2 h_z \Omega^{-3}$, and $\xi^2 J_z \Omega^{-3}$.


By performing the same procedures as in the main text, we can derive the operator form of the sub dominant terms, which are
\begin{eqnarray}
    -\int^t dt_1\int^{t_1} dt_2\int^{t_2} dt_3 \Biggl(\Bigl[[\hat{V}(t_3), \hat{H}_0], \hat{H}_0\Bigr] + \frac{1}{2}\Bigl[\hat{V}(t_2),[\hat{V}(t_3), H_0]\Bigr]\Biggr).
\end{eqnarray}
This expression corresponds to the third order Floquet-Magnus expansion.

\section{S4. \ \ \label{app:Res} Quest for the appropriate rotating frame under the resonant drive}
\begin{figure}
    \centering
    \includegraphics[width=0.4\linewidth]{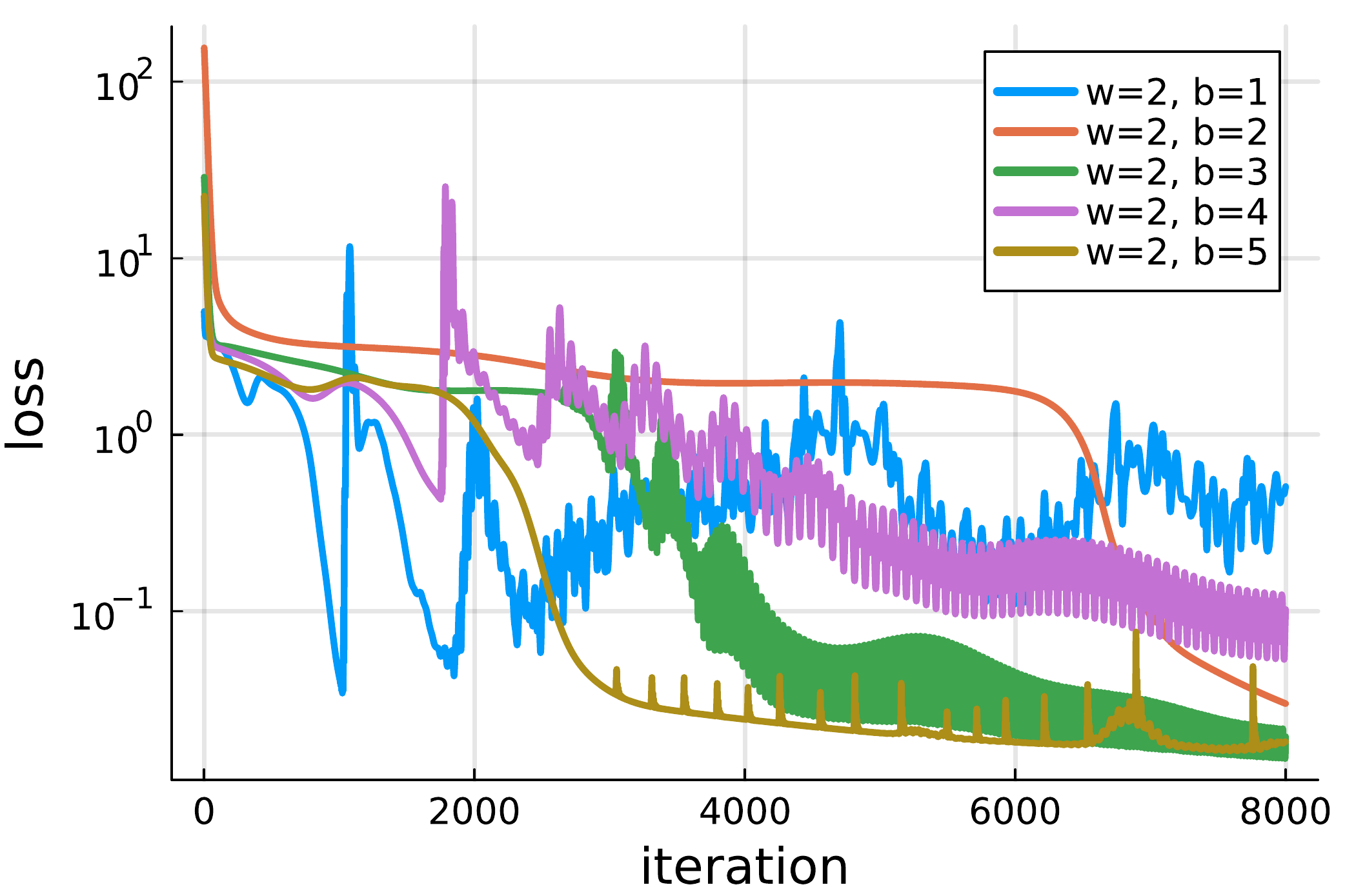}
    \includegraphics[width=0.4\linewidth]{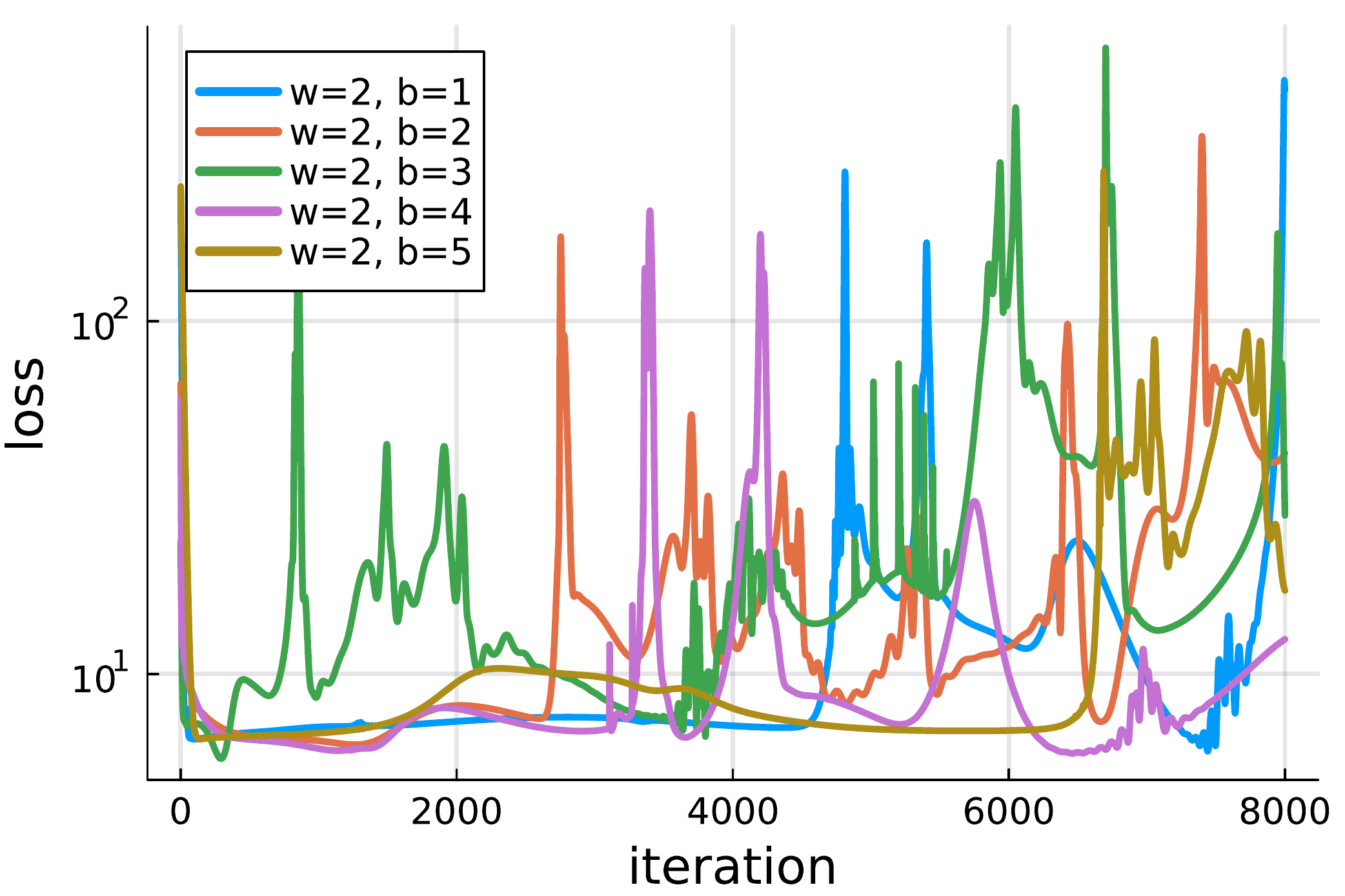}
    \includegraphics[width=0.4\linewidth]{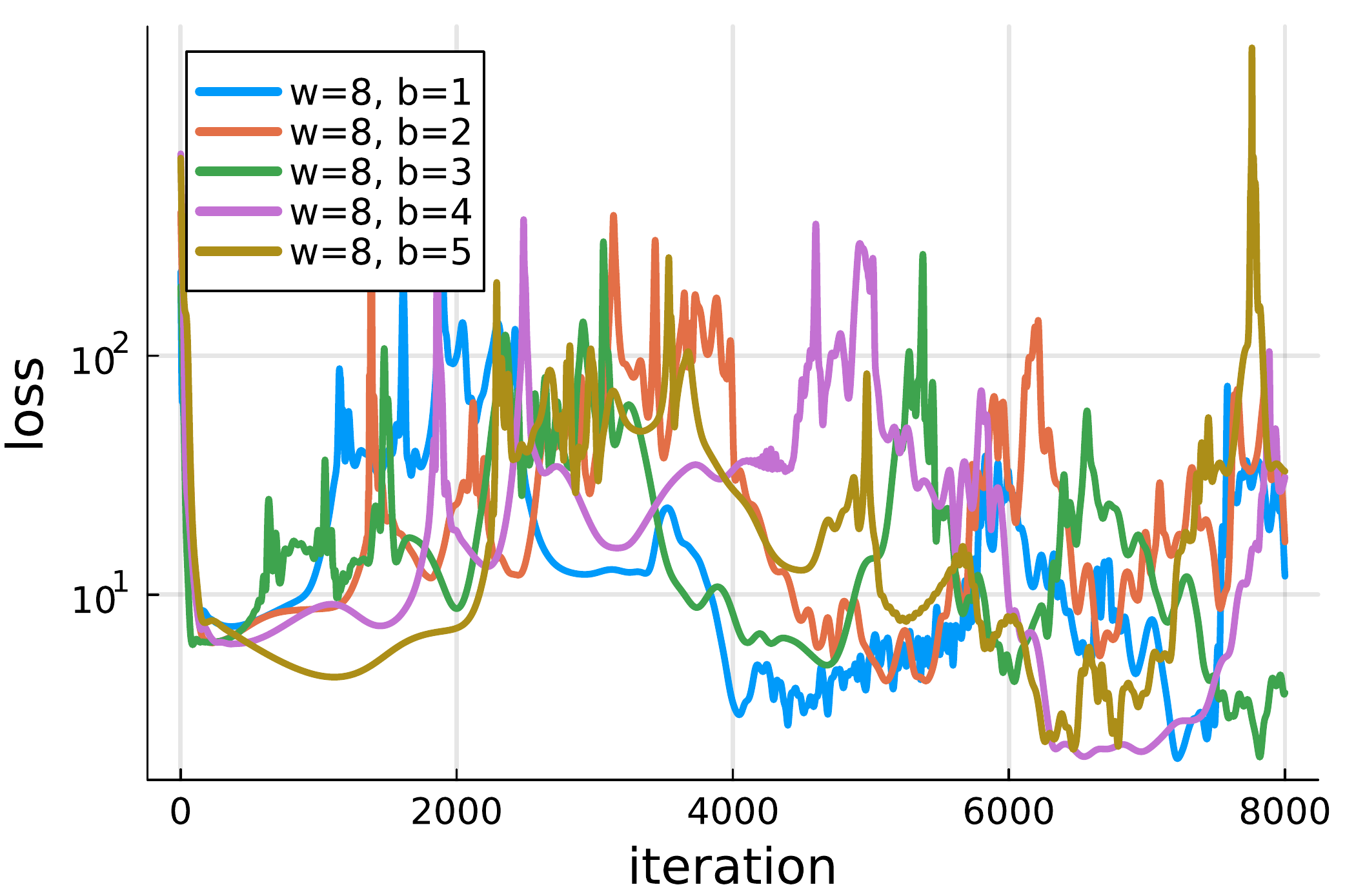}
    \includegraphics[width=0.4\linewidth]{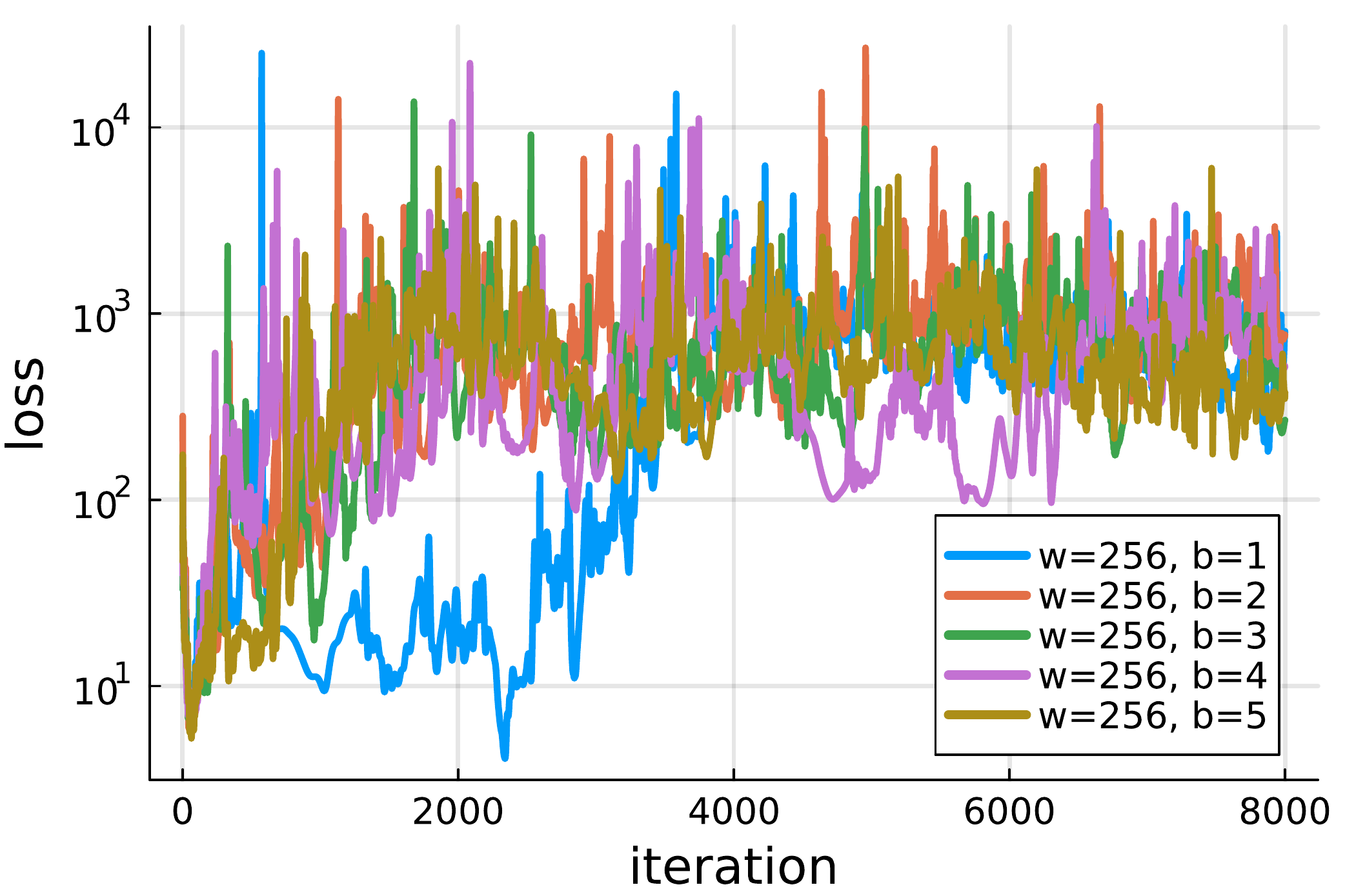}
    \caption{Hidden layer width dependence of learning dynamics in the resonant regime.$w$ represents the width of the hidden layers and $b$ represents the batch index.}
    \label{fig:res_width}
\end{figure}

In this section, we show the results of the resonant regime calculated by our proposed method. First, we check the learning dynamics and its dependence on the hidden layer width. 
Figure~\ref{fig:res_width} shows the learning dynamics at $\Omega=4.0,w=2$ and $\Omega=2.0$. We can see that the learning dynamics become unstable and rarely achieve the appropriate RF as the drive approaches resonant, and we could not achieve the appropriate RF with a small loss.  
Remembering that the RNN can achieve the appropriate RF independent of the initial condition (independent of the batch index) in figure~\ref{fig:batchdep}, these results should imply that the loss function of the parameter space has a single minimum with gentle gradient when there is a scale separation (the scale separation between $\Omega$ and the energy scale of the system), while it has some or many local minima with steep gradient when the scale separation is smeared.

\begin{figure}
    \centering
    \includegraphics[width=0.4\linewidth]{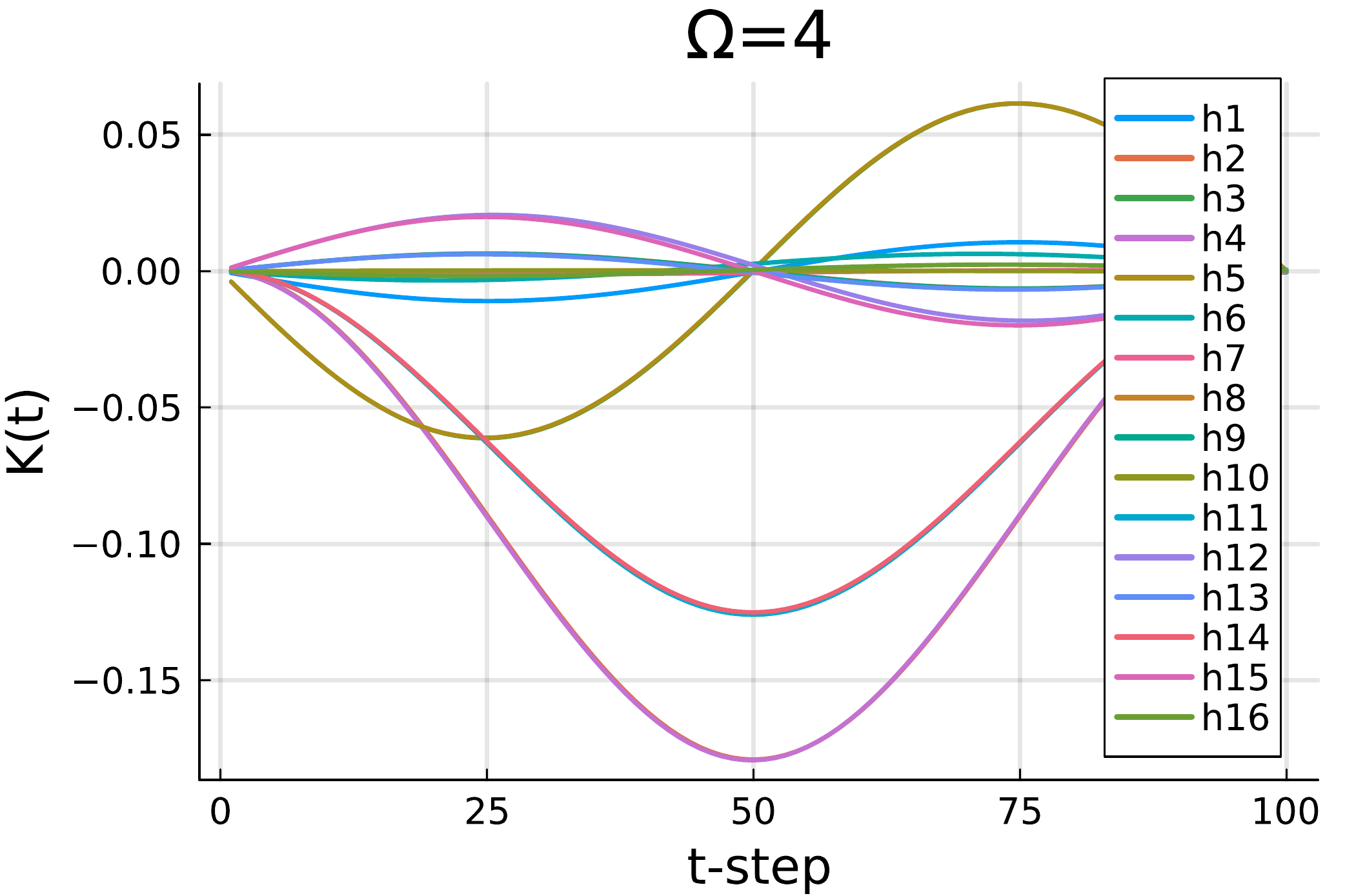}
    \includegraphics[width=0.4\linewidth]{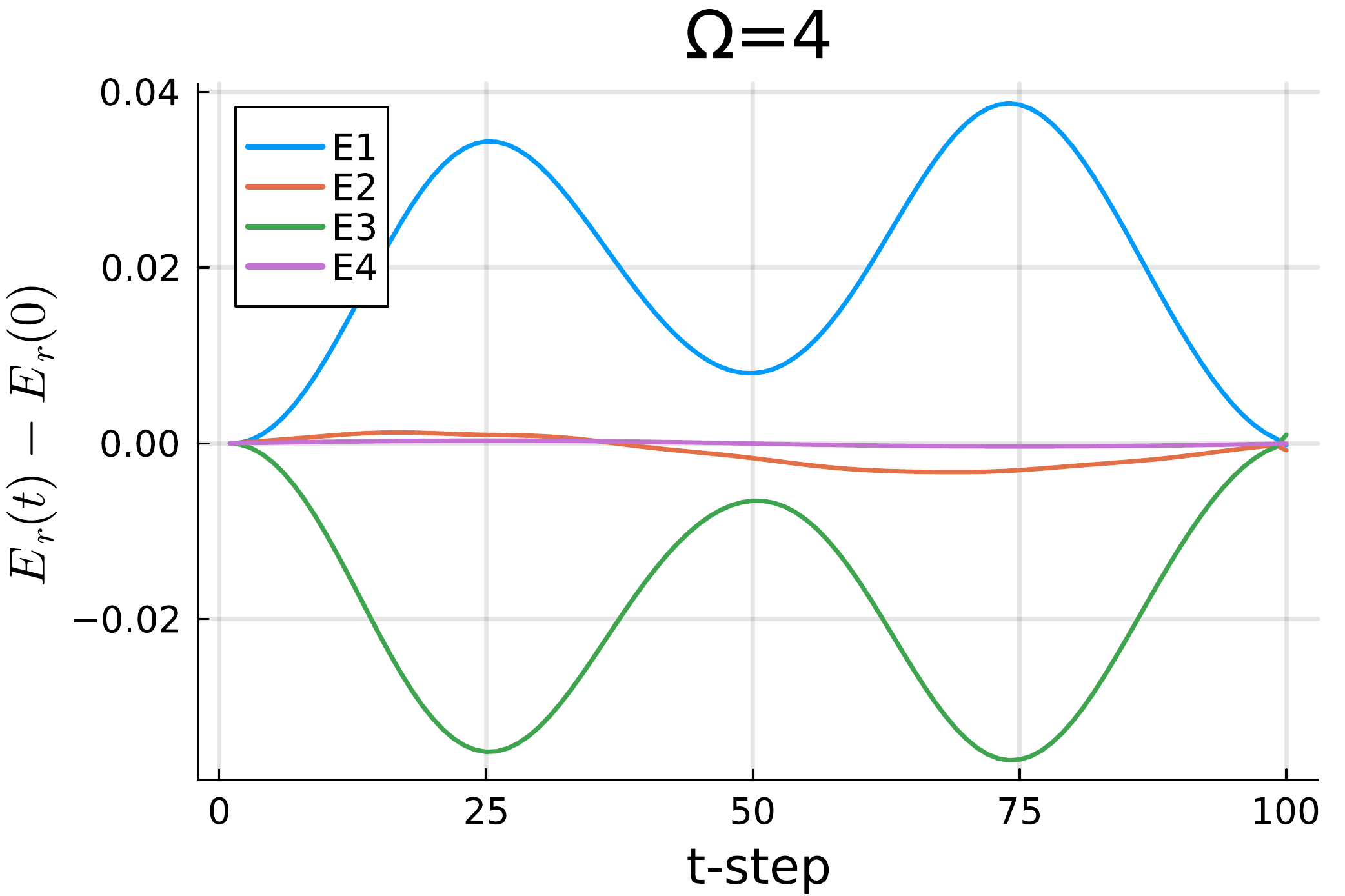}
    \caption{Best results ($w=2, b=5$) in the middle regime at $\Omega=5$}
    \label{fig:mid_Kt}
\end{figure}

\begin{figure}
    \centering
    \includegraphics[width=0.4\linewidth]{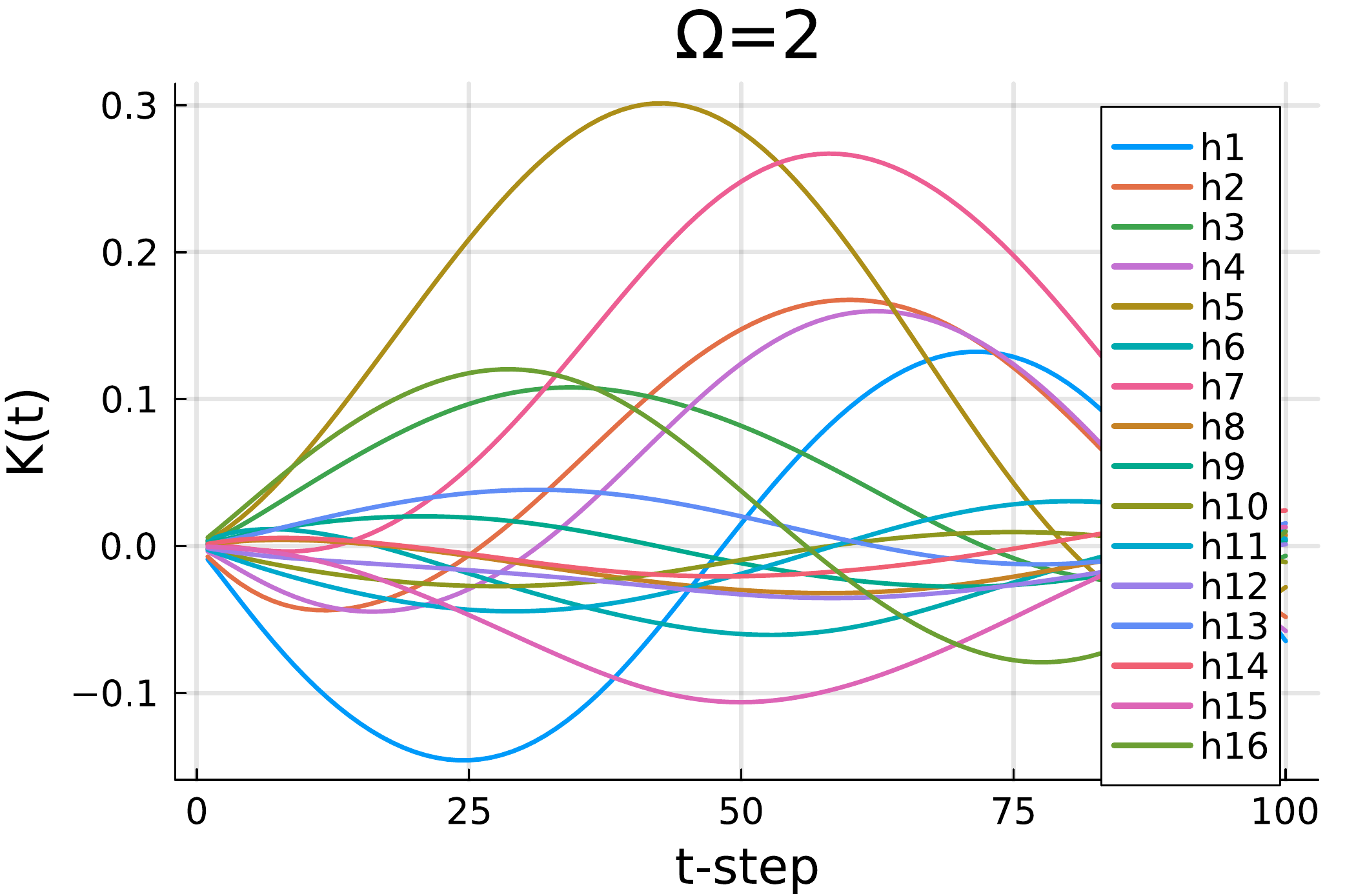}
    \includegraphics[width=0.4\linewidth]{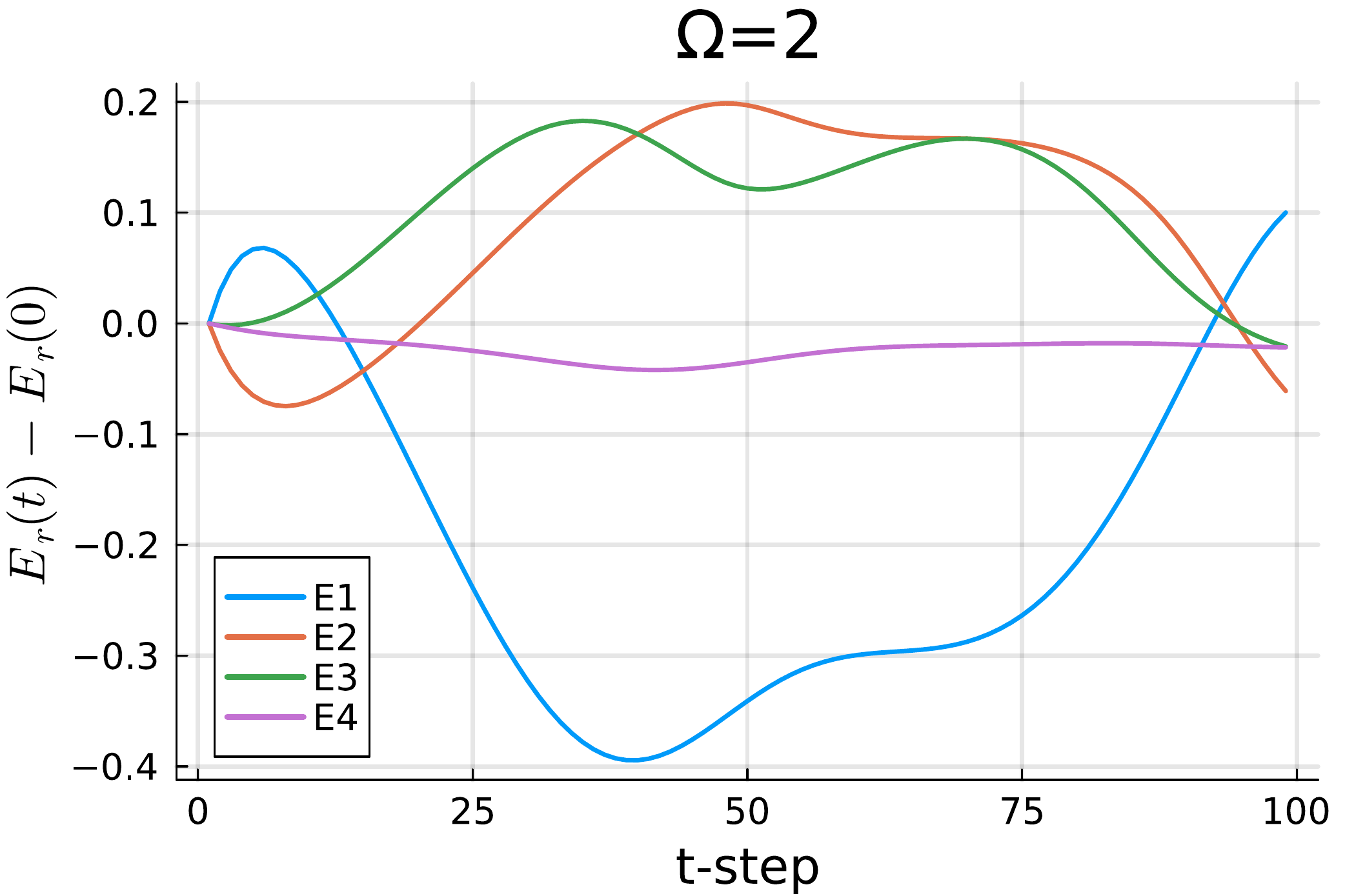}
    \caption{Best results ($w=8, b=4$) in the resonant regime at $\Omega=2$}
    \label{fig:res_Kt}
\end{figure}

We also show the best of the resulting kick operator and the dressed Hamiltonian in figure~\ref{fig:mid_Kt} and \ref{fig:res_Kt}. In the middle regime at $\Omega=4$ in figure~\ref{fig:mid_Kt}, the RNN gives us the appropriate RF which can be constructed with the high-frequency expansion (up to $1/\Omega^3$), while its learning dynamics is unstable.
In the resonant regime in figure~\ref{fig:res_Kt}, the resulting kick operator seems different from the Floquet-Magnus expansion(see and compare with figure~\ref{fig:mid_Kt}), and the resulting dressed Hamiltonian does not get the small time dependence and the time-periodicity. This failure of the quest for the appropriate RF in the resonant regime might imply that there is no ``implicit'' scale separation and a simple construction of the appropriate RF, or there are many sharp potentials for the training and it is difficult to reach the appropriate RF. The improvements in the RNN framework might lead to the discovery of the appropriate RF in the resonant regime, which is left for future works.


\end{document}